# Design and operations of load-tolerant external conjugate-T matching system for the A2 ICRH antennas at JET

I Monakhov[1], M Graham[1], T Blackman[1], S Dowson[1], F Durodie[2], P Jacquet[1],
J Lehmann[1], M-L Mayoral[1], M P S Nightingale[1], C Noble[1], H Sheikh[1], M Vrancken[2],
A Walden[1,a], A Whitehurst[1], E Wooldridge[1] and JET-EFDA contributors[b]

JET-EFDA, Culham Science Centre, Abingdon, OX14 3DB, UK
[1]EURATOM/CCFE Fusion Association, Culham Science Centre, Abingdon, Oxfordshire, OX14 3DB, UK.
[2]LPP-ERM/KMS, Association 'EURATOM-Belgian State', 1000 Brussels, Belgium

E-mail: igor.monakhov@ccfe.ac.uk

**Abstract.** A load-tolerant External Conjugate-T (ECT) impedance matching system for two A2 Ion Cyclotron Resonance Heating (ICRH) antennas has been successfully put into operation at JET. The system allows continuous injection of the RF power into plasma in the presence of strong antenna loading perturbations caused by Edge Localized Modes (ELMs). Reliable ECT performance has been demonstrated under a variety of antenna loading conditions including H-mode plasmas with Radial Outer Gaps (ROG) in the range of 4-14 cm. The high resilience to ELMs predicted during the circuit simulations has been fully confirmed experimentally. Dedicated arc detection techniques and real-time matching algorithms have been developed as a part of the ECT project. The new Advanced Wave Amplitude Comparison System (AWACS) has proven highly efficient in detection of arcs both between and during ELMs. The ECT system has allowed the delivery of up to 4 MW of RF power without trips into plasmas with Type-I ELMs. Together with the 3dB system and the ITER-Like Antenna (ILA), the ECT has brought the total RF power coupled to ELMy plasma to over 8 MW, considerably enhancing JET research capabilities. This paper provides an overview of the key design features of the ECT system and summarizes the main experimental results achieved so far.
**PACS:** 52.55.Fa, 52.50.Qt, 28.52.Cx

## 1. Introduction

Ion Cyclotron Resonance Heating (ICRH) and Current Drive (CD) is recognized as one the key techniques in tokamak fusion program [1-3]. The capability of the ICRH to provide dominant ion heating makes it an essential part of the mix of heating and CD methods required for achieving the ITER goals [4]. The ITER operational scenarios, including the non-activated start-up, rely on 20MW of ICRH power delivered to plasma; augmenting the power levels up to 40MW is also proposed for advanced stages of operation [5]. The appropriate choice of frequency and spectrum of the launched waves allow the ICRH to preferentially heat bulk ions, minority ions or electrons and to ensure spatially controllable power deposition profiles which offer a number of additional opportunities for tokamak discharge optimisation. Recent studies indicate that injection of the ICRH power could provide the means for current profile modification, generation of toroidal plasma rotation, suppression of MHD instabilities and conditioning of the tokamak first wall [6]. These circumstances give the ICRH a favourable outlook for future application in DEMO [7], where Fast Wave Current Drive (FWCD) might also play an important role.

---

[a] Sadly passed away in December 2007
[b] See appendix to Romanelli F et al 2012 *Proc.24th IAEA Conf. on Fusion Energy (San Diego 2012)* (Vienna: IAEA)

It is noteworthy that the ICRH&CD phenomena and consequently the performance of the RF heating systems are quite sensitive to the parameters of both bulk and Scape-Off Layer (SOL) plasma. In this respect JET has an advantageous position among the existing tokamaks: large-scale plasmas with significant antenna-plasma distances, variety of operational regimes including H-mode and advanced scenarios, the all-metal first wall and the RF power capabilities comparable with ITER offer the best opportunities for the assessment of reactor-relevant aspects of the ICRH physics and technology [8-11]. Since 1994 JET has been equipped with four identical A2 antennas each comprising an array of four straps [12]; new compact ITER-Like Antenna (ILA) was installed in the tokamak main port in 2008 [10]. The RF plant at JET has convenient modular design [13] capable of energizing the antennas at different frequencies with controllable strap phasing and launched power.

The practical implementation of the ICRH technique in tokamaks faces significant challenges. Among other factors, the success of the method depends on efficient and reliable operation of the RF plant delivering multi-MW power levels to phased antenna straps in the presence of small and variable plasma load [14,15]. The main difficulty in meeting this requirement stems from the fact that only a fraction of the RF power reaching the ICRH antenna is radiated into plasma while the rest is reflected back to the RF plant. In order to prevent this power from damaging expensive generator tubes, dedicated tuneable elements are introduced in the transmission lines which make the returning power circulate in a resonant circuit outside the generator end-stage. In the RF terms, this procedure implies matching the RF generator and antenna impedances. Accomplishing this task encounters a number of technical complications, e.g. mutual coupling between the antenna straps, controllable strap phasing, compatibility with operations over a band of frequencies and electrical strength limitations. By far the most serious impedance matching problem is presented by the occurrence of fast plasma events such as large 'sawtooth' crashes and Edge Localized Modes (ELMs) which quickly and strongly modify the edge plasma density. The antenna loading perturbations associated with ELMs [16-18] are characterized by particularly large magnitude (up to a 10-fold loading increase), short time-scale (~1-2 ms duration with ≤100 μs rise time) and high repetition rate (10-100 Hz) which are far beyond the capabilities of traditional impedance matching schemes relying on mechanical adjustments of the circuit elements. As a result, the harmful ELM-related power reflections trigger the RF plant protection and cause substantial reduction of the average power levels delivered to plasma. Taking into account that the ELMs are ubiquitous feature of H-mode plasmas prevailing in high-performance tokamak scenarios [19], ELM-tolerance has been identified as one of the critical issues in the development of ICRH systems in ITER and DEMO [1,7].

Several techniques have been proposed over the recent years to ensure the RF plant immunity to ELMs [14, 15]. The most common one involves installation of 3dB couplers [20-22] which divert the RF power reflected from a pair of conventional matching circuits to a dedicated dummy load. This approach has proven to be quite successful in the existing ICRH systems which fully justifies its selection for ITER [23]. However, the prospects of using this method in the next-step fusion devices are much less clear. Indeed, substantial bulkiness of the 4 MW coupler units and their infrastructure (dummy loads, coolant tanks, pumps and pipes) will dramatically increase the spatial requirements for the ~200 MW RF system in DEMO. On the other hand, the enhanced RF power losses during ELMs inherent to the method are not consistent with the expectations of high efficiency of future power plant [7]. It should also be mentioned that the $\pi/2$ phase difference between the forward voltages on the output ports of the 3dB couplers and the associated sensitivity to mutual coupling between the antenna straps [24] is not favourable for design of compact powerful multi-strap antennas in future tokamak-reactors.

An alternative ELM-tolerant impedance matching scheme known as the 'conjugate-T' capable of addressing the problems mentioned above has lately come into the focus of attention [25-27]. Substantial progress has been made in the experimental assessment of this approach in the framework of the main-port ITER-Like Antenna (ILA) projects at JET [10] and Tore Supra [28] where the in-vessel capacitors were used for matching. In parallel, the conjugate-T principle has been applied at JET to match the impedances of the existing A2 antennas using the externally located phase shifters [29]. The positive results of the initial 'proof-of-principle' assessment [30] were followed by successful commissioning of the full-scale External Conjugate-T (ECT) system. The ECT is now routinely used during ICRH operations at JET delivering into ELMy plasmas up to 4MW from two A2 antennas [11]. Together with the ITER-like antenna and the 3dB couplers installed on further pair of A2 antennas [22], this has considerably enhanced the research program at JET. The combined operation of all ELM-tolerant systems in 2009 delivered more than 8 MW into H-mode plasmas [11], increasing the confidence in successful ICRH operations in ITER.

This paper provides an overview of the key design features of the ECT system at JET and summarizes the main experimental results achieved so far. The supporting simulations are partially described in the conference presentations [29, 31]; useful theoretical treatment of a matching system similar to the ECT can also be found in a recent study [32]. The paper is organized as follows: the main aspects of application of the Conjugate-T principle in the ICRH systems are briefly reviewed in Section 2 highlighting the expected advantages and motivation for the ECT project. The basic principles of the ECT impedance matching are introduced in Section 3 where a simplified circuit is treated analytically and numerically to illustrate the main features of ECT design and operations. The details of the practical implementation of the ECT at JET are given in Section 4, including the discussions of the ECT power measurements, the issues of the operational frequency bands and the commissioning procedures. The experimental results of the ECT operations during ELMy plasmas are presented in Section 5 together with some insight into the vital ECT supplementary techniques such as the arc detection and real-time matching control. The main achievements and challenges to the injection of high power using the ECT scheme are also highlighted in this section.

## 2. Application of the Conjugate-T principle in the ICRH systems

In its general form, the conventional narrow-band impedance matching scheme [33] can be described as a parallel resonant circuit made up of two tuneable reactive elements where the small resistive load $R_0$ is connected in series to one of them (figure 1(a)). Adjustment of the reactive elements allows tuning the circuit to resonance at the operational frequency such that the impedance $Z_{T0}$ 'seen' by the generator is purely real, Im($Z_{T0}$)=0, and equals the desired value $R_{T0}$=Re($Z_{T0}$) > $R_0$. Although the considerations below refer to a simple lumped-element representation of the scheme, the conclusions remain qualitatively relevant to the case when distributed elements are used instead. In the ICRH systems, the reactances $X_{01}$ and $X_{02}$ are typically provided respectively by a line stretcher ('trombone') and a stub or by a double-stub tuner. Any perturbation of the loading resistance $R \neq R_0$ results in rapid detuning of the circuit accompanied by significant changes of both real and imaginary parts of the impedance presented to the generator, e.g. Re($Z_T$) ~ $1/R$ and Im($Z_T$) $\rightarrow -X_{02}$ during strong ELM-related loading increase $R$ (figure 1(b) and (c)). The implications of $Z_T$ deviation from $R_{T0}$ are traditionally described by the Voltage Standing Wave Ratio (VSWR) in a transmission line with the characteristic impedance $Z_0$ equal to $R_{T0}$. This parameter, denoted further as $S$, represents the ratio of the maximum and minimum voltage amplitudes in the line. It can be shown that the VSWR in the described circuit is linearly proportional to the perturbed loading resistance when $R$>$R_0$ (figure 1(d)) indicating that the conventional matching scheme is poorly compatible with ELMs.

The Conjugate-T impedance matching, discussed in this paper, relies on the same principle of tuneable parallel resonance; however, the fundamental difference is that the resistive loads are included in both branches of the circuit and the reactances $X_{01}$ and $X_{02}$ have equal absolute values $X_{01}$=$X_{02}$=$X_0$ (figure 2(a)). These features dramatically change the circuit's response to loading perturbations providing the perturbations are identical in both the branches. Indeed, the impedance $Z_T$ 'seen' by the generator now remains purely real, Im($Z_T$)=0, regardless of the loading resistance $R$ (figure 2(c)) and the real part Re($Z_T$) does not monotonically decrease at high $R$ but starts to rise slowly and reaches $R_{T0}$ again (figure 2(b)). The resulting VSWR dependence (figure 2(d)) turns to be much 'flatter' and closer to unity over a wide range of the loading resistances. This behaviour explains high ELM-tolerance of the Conjugate-T scheme.

It should be noted that the aforementioned favourable properties have practical significance only when the matched (unperturbed) resistances $R_0$ and $R_{T0}$ are of the same order; otherwise the minimum value of Re($Z_T$) deviates from $R_{T0}$ too much and the corresponding local maximum of the VSWR gets unacceptably high. This complicates the application of Conjugate-T scheme in the ICRH systems where the loading resistance $R_0$ is typically much smaller than the output resistance of the RF generator $Z_0$ and direct matching of $R_0$ and $Z_0$ (i.e. such that $R_{T0}$=$Z_0$) does not ensure sufficient ELM-tolerance. The solution of this problem requires two-stage matching where tuneable Conjugate-T circuit is used to match $R_0$ to an intermediate resistance $R_{T0}$ ($Z_0$>>$R_{T0}$>$R_0$) and a complementary second-stage circuit provides the final transformation of $R_{T0}$ to $Z_0$ (figure 3(a)). Thus, an ICRH system based on Conjugate-T principle generally comprises two antenna straps equipped with variable reactive elements which are tuned to form a pair of complex-conjugate impedances; the impedances are connected in parallel at a T-junction and the resulting impedance is further transformed by a dedicated second-stage circuit to match the output impedance of the generator.

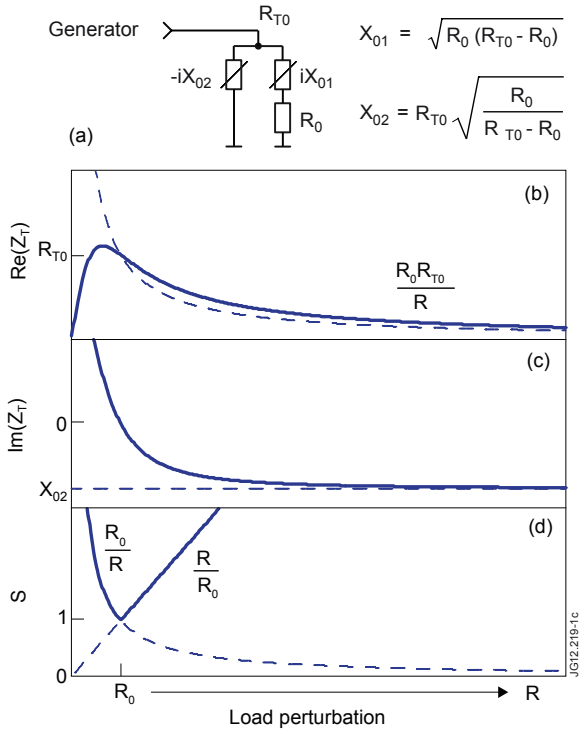
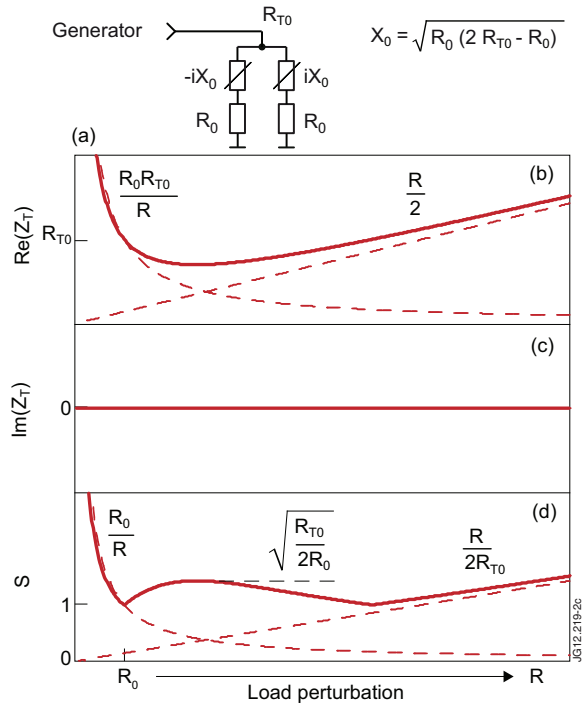

**Figure 1.** Generalized conventional impedance matching scheme (a) and its response to the loading resistance perturbation: real (b) and imaginary (c) parts of the impedance seen by the RF generator and the VSWR in the generator output transmission line (d). The dashed lines show asymptotic estimates produced in assumption of $R_0 \ll R_{T0}$.

**Figure 2.** Generalized conjugate-T impedance matching scheme (a) and its response to the loading resistance perturbation: real (b) and imaginary (c) parts of the impedance seen by the RF generator and the VSWR in the generator output transmission line (d). The dashed lines show asymptotic estimates produced in assumption of $R_0 \ll R_{T0}$.

As an ELM-tolerant technique, the Conjugate-T matching offers several additional advantages over the alternative methods. Unlike the case of the 3dB couplers, the RF power is delivered to plasma during the whole duration of an ELM thus increasing the system efficiency. The phase difference $\Delta\varphi$ between the voltages on the conjugated antenna straps, $\pi \geq \Delta\varphi > \pi/2$, (see Section 4.1) better suits closely packed antenna arrays than $\Delta\varphi = \pi/2$ phasing typical of the quadrature hybrids: the poloidal radiation spectrum could be made more favourable to antenna-plasma coupling and adverse effects of mutual coupling between the straps are less pronounced. Crucially, the Conjugate-T approach doesn't require dedicated bulky infrastructure allowing a compact design.

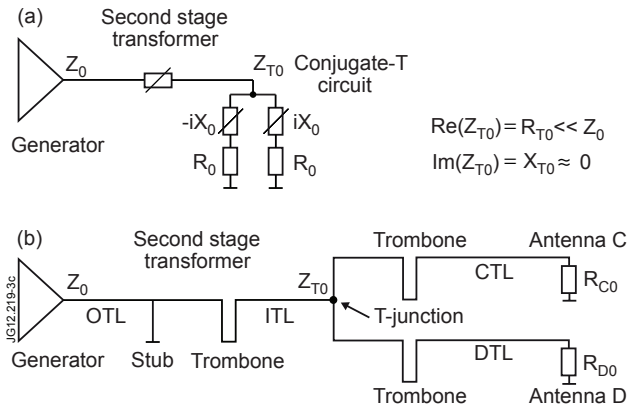

**Figure 3.** Generalized matching scheme based on the Conjugate-T principle (a) and the External Conjugate-T circuit configuration adopted at JET (b).

The idea of using the matching scheme where resistive loads are present in both branches of tuneable parallel resonant circuit (figure 2(a)) emerged in the 1980s when Resonant Double Loop (RDL) ICRH launcher design was proposed [34]. The focus of the RDL design was a compact high-power main-port antenna directly matched to the generator output impedance $Z_0$ within the tokamak vacuum boundary, minimizing voltages at the vacuum feedthroughs and reducing power losses in the mismatched part of the system. This particular framework did not allow the intrinsic load tolerance of the scheme to be fully revealed in early RDL proposals for TFTR, Tore Supra [35] and DIII-D [16]. Later realization of the significance of Conjugate-T matching to an intermediate resistance of $R_0 \ll Z_0$ [26,27] stimulated further development of the RDL concept during the design of ELM-tolerant ICRH system on ITER [36]. In line with the original considerations, the advanced RDL design relied on in-vessel reactive elements closely integrated into the ICRH antenna structures. Due to tight space restrictions, lumped vacuum capacitors appeared to be the most promising option for building the tuneable conjugate-T circuits.

Substantial efforts were committed to the experimental assessment of the in-vessel conjugate-T matching within the ITER-Like Antenna (ILA) projects at JET [10] and Tore Supra [28]. The projects have proven the viability of the method confirming enhanced load tolerance of the conjugate-T scheme. At the same time, the experience of practical design, commissioning and operations of the ILA systems has revealed a number of shortcomings of the adopted approach to the implementation of the conjugate-T principle. The in-vessel location of the tuning elements significantly complicates solution of several reactor-relevant antenna design problems, such as the reparability, cooling and mechanical stress management; the introduction of mechanical drive system into the vacuum vessel provides strong technical challenges and, eventually, reduces the reliability of the system. The essential requirement of controllable phasing of the toroidally adjacent antenna straps limits the in-vessel matching to the case where the conjugated straps are arranged poloidally. In this situation, peculiarities of poloidal plasma shape inevitably bring an asymmetry in the strap design and plasma loading (see e.g. [37]) which has negative impact on the system performance. Further unfavourable asymmetries are introduced by the mutual coupling between the conjugated straps due to their close proximity. Insufficient transparency of the scheme to conventional remote RF diagnostics such as directional couplers necessitates installation of vulnerable in-vessel sensors e.g. voltage probes or pickup coils [38] that can be very difficult to repair in reactor conditions. The installation of in-vessel sensors complicates the antenna vacuum interfaces and the complexity of the sensor in-situ calibration contributes to difficulties with the circuit control and analysis. Finally, the expectations of enhanced electrical strength of the antenna with in-vessel matching appear to be burdened by a new challenging problem related to the electrical vacuum breakdown at the T-junction [39] which requires development of sophisticated dedicated arc detection techniques [40].

Structural separation of the ICRH antenna and its matching system and the external (i.e. outside the vacuum vessel) positioning of the variable matching elements removes most of the complications mentioned above [25,29,41]. Such an approach allows the use of simple technical solutions substantially reducing the cost and bringing new possibilities to the RF plant design; in particular, it makes the conjugate-T principle applicable to a variety of antenna configurations including the pre-existing ones [25]. Importantly, the external location of the matching elements removes the requirement for close spatial proximity of the conjugated straps. Thus, the straps belonging to separate antennas could be paired [29] which provides an ultimate solution to the problem of the strap mutual coupling, plasma coupling asymmetries and controllable antenna strap phasing.

The choice of appropriate variable matching elements is a key issue in implementation of the external conjugate-T scheme. Tests of the systems based on variable capacitors [42] and stubs [43] have been reported, however, reliance on such elements in large-scale broadband applications is fraught with difficulties. Indeed, the global characteristics of such circuits are critically sensitive not only to the reactances introduced by the elements into the lines connecting the T-junction and the antenna straps but also to the absolute position of the elements along these lines. This circumstance strongly affects the frequency response, complicates the control and optimisation and puts constraints on the installation of the system. Using line stretchers (trombones) as external matching elements [25,29,41], significantly simplifies the circuit behaviour and gives a number of practical advantages. Unlike the capacitors and stubs, the trombones don't affect the amplitude of the line voltage waves and their exact physical location in the lines is of no consequence. At the same time, they provide a straightforward means of compensation of the line electrical length dispersion over the frequency band and easily accommodate the antenna strap equivalent electrical length change in the presence of plasma. The trombone-based scheme is highly transparent for analysis and diagnostics including the antenna monitoring; a simple set of directional couplers with arbitrary location in the lines is capable of providing adequate and easily interpretable information about the critical system parameters.

Another distinctive feature developed within the conjugate-T projects at JET is related to the strategy of using the variable impedance transformer [29,44], specifically an adjustable stub-trombone tuner, for the second-stage matching (see figure 3(b)). As mentioned earlier, the primary purpose of the transformer is matching the small intermediate impedance $Z_{T0}$ at the T-junction and the output impedance of the generator $Z_0$. Initially, fixed multistage quarter-wavelength transformers or Klopfenstein tapers were considered for this task [41], however, compatibility with broad-band operations required bulky design relying on non-standard transmission line components. The stub-trombone approach eliminated this problem and brought new qualities to the system. The transformer adjustability allows the ability to turn the T-junction impedance into a useful additional 'control handle' [45], giving the whole scheme substantial flexibility while keeping the generator and antenna impedances matched. The control over the real part of the T-junction impedance helps to optimize the system tolerance to the resistive component of the antenna loading perturbation during ELMs in different plasma scenarios. Adjustments of the imaginary part of $Z_{T0}$ make it possible to optimize the response to the antenna reactance perturbations and, importantly, to equalize the voltages on the straps in the presence of loading asymmetries. The latter circumstance is vital for achieving the maximum power performance of the ICRH system.

The above-mentioned expectations for efficient ELM-tolerant ICRH system alternative to the 3dB couplers have been realized at JET in the form of the External Conjugate-T (ECT) project [31] for the A2 antennas [12]. The project takes advantage of many of the favourable opportunities discussed above, including pairing the corresponding straps belonging to different ICRH antennas, relying on the trombones as the matching elements and using the adjustable stub-trombone tuners for the second-stage impedance transformation. The in-depth description of the ECT design and the main experimental results are given below.

## 3. Basic principles of the ECT design and operation

*3.1 Simplified ECT scheme and the principles of impedance matching*

The detailed discussion of the ECT system requires the introduction of some basic features and notations using a simplified scheme shown on figure 3(b). Two antenna straps are represented here by the equivalent resistances $R_{C0}$ and $R_{D0}$ accounting for the RF power dissipation due to radiation into plasma and Ohmic losses; the subscripts 'C' and 'D' correspond to the naming conventions used for the ICRH antennas involved in the actual ECT installation on JET. The unperturbed ('between-ELM') equivalent strap resistances will be denoted by the index '0' in this paper. In conditions typical of JET A2 antennas, $R_{C0} \approx R_{D0} \approx 1-2\,\text{Ohm}$, while during the ELMs the resistances could increase up to $R_C \approx R_D \geq 8\,\text{Ohm}$ [17]. The straps 'C' and 'D' are connected to the coaxial T-junction by the Transmission Lines called respectively CTL and DTL. The lengths of both lines could be adjusted by the line stretchers ('trombones'). For convenience, the antenna strap reactances are treated as notional extra sections of transmission lines added to the CTL and DTL. The equivalent length of these sections depends on the operational frequency and particular strap design; in the presence of quasi-stationary plasma the equivalent lengths typically shorten by $0.1-0.15\,\text{m}$ compared with the vacuum value and during ELMs further momentary decrease in length up to $0.3-0.4\,\text{m}$ could take place [17]. The total CTL and DTL lengths including the trombones and the equivalent strap lengths are denoted as $L_{\text{CTL}}$ and $L_{\text{DTL}}$. The coaxial T-junction is connected to the generator Output Transmission Line (OTL) via a variable stub-trombone tuner which serves as the second-stage Impedance Transformer (IT). The IT transmission line between the T-junction and the stub is known as the ITL and its total length including the trombone is referred to as $L_{\text{ITL}}$. For simplicity and without loss of generality, the analytical and numerical treatment of the line lengths in this paper is performed in the $\lambda/2\pm\lambda/4$ domain, where $\lambda$ is the wavelength; the implications of the choice of the physical CTL, DTL and ITL lengths will be discussed in Section 4.3. All the transmission lines in the system have identical characteristic impedance $Z_0 = 30\,\text{Ohm}$ equal to the output impedance of the generator.

Adjustments of the CTL and DTL lengths by the trombones allow control of the parallel impedances presented by the loaded lines to the T-junction at a given operational frequency. In a practically important case, these impedances could be tuned to form a complex-conjugate pair such that the resulting impedance at the T-junction $Z_{T0} = R_{T0} + iX_{T0}$ has a predominantly resistive component which is much smaller than the characteristic impedance of the line $X_{T0} \ll R_{T0} \ll Z_0$. (Note, that the index '0' is used here to refer to the reference value of the T-junction impedance $Z_{T0}$, i.e. the value which ensures that the unperturbed 'between-ELM' strap resistances and the generator output impedance are matched in the presence of the second-stage impedance transformer; it should be recognized that the actual value $Z_T$ of the T-junction impedance seen by the ITL deviates from the reference value $Z_{T0}$ when the system is mismatched, e.g. during ELMs or if the CTL and DTL trombone lengths are not set correctly). The signs of the imaginary parts of the complex-conjugate pair of parallel impedances which form the $Z_{T0}$ are interchangeable and the desired result could be achieved with the CTL and DTL representing the inductive and capacitive branches or vice versa. The two possibilities corresponding to the $L_{CTL} > \lambda/2 > L_{DTL}$ and $L_{DTL} > \lambda/2 > L_{CTL}$ will be further identified as the Conjugate-T matching options $CT=0$ and $CT=1$ respectively (see also figure 10 in Section 4.4). In the ideal case of equal loading of the antenna straps $R_{C0} = R_{D0} \equiv R_0$ and without mutual coupling between them, the T-junction reference impedance could be optimally set to a purely real value $Z_{T0} = R_{T0}$, $X_{T0} = 0$. In this situation, the choice of the CT option makes no difference to the ECT performance, and the CTL and DTL matching lengths could be found from the following expressions:

$$L_{CTL} = \frac{\lambda}{2} \pm \frac{\lambda}{2\pi} \arctan\left(\sqrt{\frac{R_0(2R_{T0} - R_0)}{Z_0^2 - 2R_{T0}R_0}}\right) \quad \text{and} \quad L_{CTL} + L_{DTL} = \lambda \tag{1}$$

where the alternative signs correspond to the $CT=0$ and $CT=1$ matching options. In the more general case of asymmetric loading and finite mutual coupling between the antenna straps, optimisation of the ECT response to ELMs and minimisation of the strap voltages require setting the T-junction impedance to a complex value $X_{T0} \neq 0$. Under these circumstances the choice of the CT option noticeably affects the ECT behaviour [31]: the lengths no longer form an interchangeable pair of values for the two CT options and numerical methods are needed for finding the CTL and DTL matching lengths. In order to ensure maximum flexibility, the ECT hardware and control electronics at JET have been designed to be compatible with both CT matching options.

The second-stage impedance transformation $Z_{T0} \to Z_0$ is achieved by setting the ITL stub and trombone lengths to appropriate values depending on the operational frequency and the $Z_{T0}$ value. In the simple case of purely real T-junction impedance $Z_{T0} = R_{T0}$, $X_{T0} = 0$ the required lengths satisfy the following equations:

$$L_{stub} = \frac{\lambda}{2} \pm \frac{\lambda}{2\pi} \arctan\left(\frac{\sqrt{R_{T0}Z_0}}{Z_0 - R_{T0}}\right) \quad \text{and} \quad L_{ITL} = \frac{\lambda}{2} \mp \frac{\lambda}{2\pi} \arctan\left(\sqrt{\frac{R_{T0}}{Z_0}}\right) \tag{2}$$

The alternative signs in these expressions correspond to two options for setting the IT elements' lengths allowing the necessary transformation. These options, which exist for an arbitrary complex value of $Z_{T0}$, will be further referred to as the IT options $IT=0$ and $IT=1$ respectively. The choice of the IT option does not affect the ECT performance at a given operational frequency; however, the possibility of running the system in two IT configurations expands the range of the frequencies where the transformation could be accomplished in the presence of realistic stub and trombone length limitations (see Section 4.3).

In conditions of variable antenna loading, the ECT line lengths have to be adjusted in order to maintain the perfect impedance matching. Practically, this is achieved by an automatic real-time control of the CTL and DTL trombone lengths (see Section 5.3) such that the running value of the impedance at the T-junction $Z_T$ is kept close to the preferred reference value $Z_{T0}$. At the same time, the lengths of the ITL trombone and stub are kept fixed during the plasma pulse; they are changed only between the pulses if $Z_{T0}$ optimisation is needed. This approach allows the maintaining of the generator impedance match without modifying the T-junction impedance which is essential for predictable ECT performance. In principle, the two-stage impedance transformation makes $Z_{T0}$ a 'free' parameter in the matching process: technically, it could be set to an arbitrary value, including $Z_{T0}=Z_0$ thus making the second-stage transformer redundant. However, the choice of the $Z_{T0}$ is crucially important in optimising some other critical aspects of the ECT behaviour, in particular its response to ELMs.

*3.2 The ELM tolerance of the ECT scheme*

During fast antenna loading perturbations, e.g. ELMs, the inertial matching elements (trombones) cannot keep the impedance at the T-junction close to the reference value $Z_{T0}$ and the impedance seen by the generator deviates from $Z_0$. Traditionally, the OTL Voltage Standing Wave Ratio (VSWR) is used as a quantitative measure of the impedance mismatch

$$S = \frac{\left|V_{OTL}^{for}\right| + \left|V_{OTL}^{ref}\right|}{\left|V_{OTL}^{for}\right| - \left|V_{OTL}^{ref}\right|} \quad (3)$$

where $\left|V_{OTL}^{for}\right|$ and $\left|V_{OTL}^{ref}\right|$ are the amplitudes of forward and reflected voltage waves in the OTL. A value of $S<1.5$-$2$ is considered safe for full-power end-stage tetrode operation in conditions of arbitrary OTL reflection phase [23]; historically, slightly wider margin of $S<3$ is routinely used at JET [13].

The impact of ELMs on the ECT impedance matching can be illustrated using analytical expressions relevant to the simple case of $Z_{T0} = R_{T0}$, $X_{T0} = 0$ considered above (1-2). The OTL VSWR behaviour during the instantaneous symmetric strap resistance increase from $R_{C0} = R_{D0} \equiv R_0$ to $R_C = R_D \equiv R$ can be described by the following equation:

$$S = \frac{2RR_{T0}(Z_0^2 - R_0^2)}{(R^2 - R_0^2)Z_0^2 + 2R_0 R_{T0}(Z_0^2 - R^2)}, \quad \text{if } R > R_0 \quad (4)$$

where in practice the conditions of $R_0<<Z_0$ and $R<<Z_0$ allow this to be further simplified to

$$S \approx \frac{2RR_{T0}}{R^2 - R_0^2 + 2R_0 R_{T0}}, \quad \text{if } R > R_0 \quad (5)$$

(Note, that in the case of the momentary reduction of the strap resistance $R<R_0$ the expressions represent the inverted value of $S$). The formulae show that the dependence of the OTL VSWR on the antenna strap resistance $R$ during ELMs is not monotonic (see also figure 4). Together with the trivial $R=R_0$ matching condition between ELMs, the perfect $S=1$ impedance transformation is also achieved when the strap resistance reaches

$$R_{Smin} \approx 2R_{T0} - R_0 \quad (6)$$

while at certain strap resistance between the two matching points $R_0$ and $R_{Smin}$

$$R_{Smax} \approx \sqrt{R_0(2R_{T0} - R_0)} = \sqrt{R_0 R_{Smin}} \quad (7)$$

the OTL VSWR has a local maximum

$$S_{max} \approx \frac{R_{T0}}{\sqrt{R_0(2R_{T0} - R_0)}} = \frac{R_{T0}}{\sqrt{R_0 R_{Smin}}} = \frac{R_{T0}}{R_{Smax}} \quad (8)$$

The last expression is consistent with the simple estimate for $S_{max}$ shown on figure 2(d) which is based on the assumption of $R_0<<R_{T0}$. The described OTL VSWR behaviour forms the basis of the ELM-tolerance of the conjugate-T matching systems. As follows from (6-8), the optimisation of the ELM-tolerance (i.e. maximising the $R_{Smin}$ while keeping the $S_{max}$ below the allowed limit) imposes clear constraints on the $R_{T0}$ value. In conditions of JET A2 antenna strap loading $R_0 \approx 1-2\ \text{Ohm}$, the optimum settings for the reference resistance at the T-junction correspond to $R_{T0} \approx 4 \pm 1\ \text{Ohm}$.

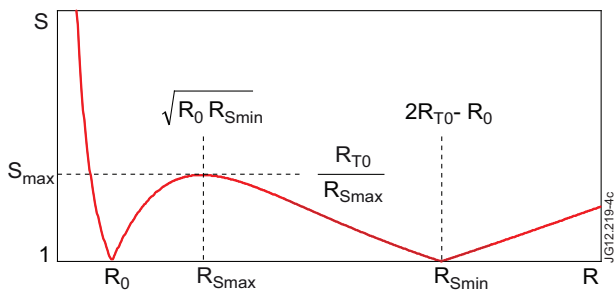

**Figure 4.** Qualitative dependence of the OTL VSWR on the antenna strap resistance for a Conjugate-T matching circuit where $R_0$ is the unperturbed (between-ELM) strap resistance and $R_{T0}$=Re($Z_{T0}$) is the real part of the reference impedance $Z_{T0}$ at the T-junction. An idealized case of equal loading of the conjugated antenna straps and purely real T-junction reference impedance Im($Z_{T0}$)=0 is illustrated; the antenna strap reactance perturbation and mutual coupling is ignored.

*3.3 The ECT impedance matching optimisation in realistic loading conditions*

The above treatment of the symmetrically loaded $R_{C0}=R_{D0}$ ECT circuit assumes that the reference impedance at the T-junction has no imaginary component $X_{T0} \equiv \text{Im}(Z_{T0}) = 0$ and it implies that the choice of the CT matching option makes no difference. The in-depth analysis shows, however, that the ECT performance optimisation may require using non-zero $X_{T0}$ values and discriminating between the CT matching options. Several factors, such as the antenna strap inductance perturbation, loading asymmetry and mutual coupling could affect the ECT behaviour in real conditions, while the optimisation criteria include not only the issues of ELM-tolerance but also the system power capabilities between ELMs. The considerations below highlight some of the key aspects of the ECT matching lying beyond the idealized case discussed so far.

In the presence of plasma, the antenna strap resistance variation is usually accompanied by strap inductance change which equally contributes to the impedance matching problem [16,17]. The inductance change could be conveniently expressed in terms of the equivalent change of the CTL or DTL length $\delta L$; a linear relationship $\delta L \approx k_L \delta R$ is typically observed between the A2 antenna strap resistance perturbation $\delta R = R - R_0$ and $\delta L$ where the proportionality constant $k_L$ has a negative value between $-1\,\text{cm/Ohm}$ and $-5\,\text{cm/Ohm}$ depending on the plasma scenario [17]. The rapid equivalent line length change during ELMs could have noticeable impact on the ELM-tolerance requiring dedicated circuit optimisation. Figure 5 shows the results of simulations of the OTL VSWR response to complex antenna loading perturbation at $f$=42.5 MHz in the absence and in the presence of the strap loading asymmetry. Figures 5(a-b) correspond to the case of identical loading $R_{C0}=R_{D0}$=1 Ohm of both antenna straps and different imaginary parts of the T-junction reference impedance: (a) $X_{T0}$=0 Ohm and (b) $X_{T0}$=-1 Ohm. These two plots are not sensitive to the choice of the CT matching option. Figures 5(c-d) correspond to the case of $X_{T0}$=-0.5 Ohm and asymmetric strap loading $R_{C0}$=1.2 Ohm and $R_{D0}$=0.8 Ohm when the OTL VSWR behaviour depends on the choice of the CT option: (c) $CT$=0 and (d) $CT$=1. The horizontal axes of the plots represent relative variation of the strap loading resistances with respect to their unperturbed ('between-ELM') values $R_{C0}$ and $R_{D0}$. The strap equivalent length change is assumed to be linearly proportional to the strap resistance deviation from the unperturbed values; the proportionality constant $k_L$ varies along the vertical axes of the contour plots; the plots at the bottom of the contour plots show the VSWR 'slices' at $k_L$=-3 cm/Ohm. The real part of the T-junction reference impedance used in the simulations was $\text{Re}(Z_{T0}) \equiv R_{T0} = 3\,\text{Ohm}$. It follows from the plots that the antenna reactance perturbation affects the ELM tolerance of the circuit (figures 5(a-b)); to achieve the best results the reference reactance at the T-junction must be set to $X_{T0}$<0. Loading asymmetries between the conjugated straps further complicate the circuit response to ELMs and, depending on which strap experiences higher loading, discrimination between the CT matching options is needed (figures 5(c-d)): in conditions of $X_{T0}$<0 it is beneficial to have the strap with lower resistance in the inductive branch of the circuit.

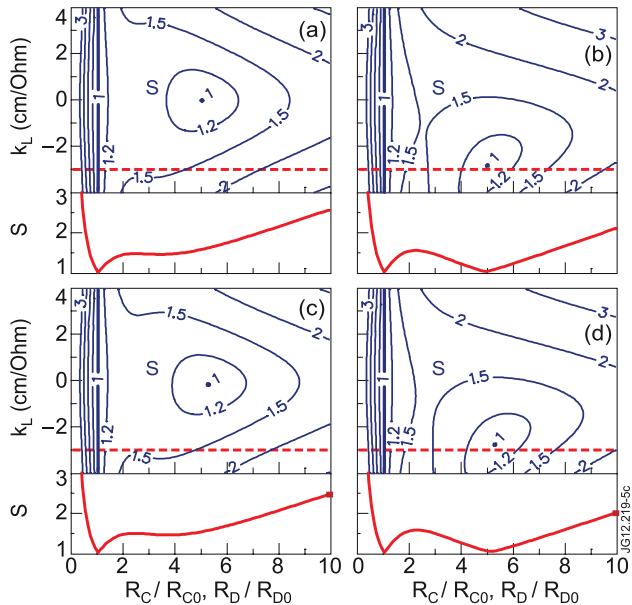

**Figure 5.** Simulations of the OTL VSWR response to complex antenna loading perturbation during ELMs. The plots (a-b) correspond to the case of $R_{C0}=R_{D0}$=1 Ohm and different imaginary parts of the T-junction reference impedance: (a) $X_{T0}$=0 Ohm and (b) $X_{T0}$=-1 Ohm. The plots (c-d) correspond to the case of $X_{T0}$=-0.5 Ohm and asymmetric strap loading $R_{C0}$=1.2 Ohm, $R_{D0}$=0.8 Ohm when the OTL VSWR behaviour depends on the choice of the CT option: (c) $CT$=0 and (d) $CT$=1. See section 3.3 for detailed explanation of the plot features.

The loading issues discussed above were addressed during the ECT design. One of the major complicating factors, specifically the mutual coupling between the conjugated straps, was entirely eliminated in the ECT implementation adopted at JET where the conjugated straps belong to separate antennas. Other measures included pairing the straps with identical design and position within the antenna arrays which helped to minimize the loading asymmetries. At the same time, possible differences between the local plasma density profiles in the scrape-off-layer in front of the antennas still leave this issue relevant. The adjustability of the second-stage impedance transformer and the capability to choose between the CT options provided in the ECT design are valuable additional tools for dealing with the problem: the characteristics of the circuit can be modified between plasma pulses to accommodate a particular loading behaviour and to optimize the ELM-tolerance.

It should be noted that the technique of adjusting of the imaginary part of the T-junction impedance has clear limitations related to maximising the ECT power performance between ELMs. Indeed, both the $X_{T0}$ deviation from zero and the strap loading asymmetry, $R_{C0}/R_{D0} \neq 1$, lead to an imbalance between the maximum voltages in the conjugated lines and the condition of $\left|V_{CTL}^{max}\right| = \left|V_{DTL}^{max}\right|$ is fulfilled only if certain relationship between the $X_{T0}$ and $R_{C0}/R_{D0}$ is maintained depending on the CT matching option. This behaviour is illustrated on figure 6(a) where contour plots show the ratios of the maximum voltages in the conjugated lines $\left|V_{CTL}^{max}\right|/\left|V_{DTL}^{max}\right|$ calculated for the matching option $CT$=0 (solid lines) and $CT$=1 (dashed lines) for a range of values of $X_{T0}$ and $R_{C0}/R_{D0}$. The simulations were performed assuming $f$=42.5 MHz and $R_{T0}$=3 Ohm and the average strap resistance was kept constant $(R_{C0} + R_{D0})/2 = 1$ Ohm during variation of the $R_{C0}/R_{D0}$ ratio. The voltage imbalance reduces the maximum power deliverable by the ECT system as the electrical breakdown limitations in one strap are reached while the second strap still has not used its full power capability. Assuming that both antenna straps have identical electrical strength, the reduction of the ECT total power capability due to the voltage imbalance is illustrated by figure 6(b). The system utilisation factor presented on this plot is defined as a percentage of the maximum deliverable power which is coupled to plasma when the maximum voltage limit is reached in one of the lines. It follows from figure 6(b), that delivering the maximum power between ELMs does not allow substantial deviation of $X_{T0}$ from zero (typically, $-0.5$ Ohm $< X_{T0} < 0$ Ohm) which means that a compromise with the optimum ELM-tolerance requirements may be needed in practice.

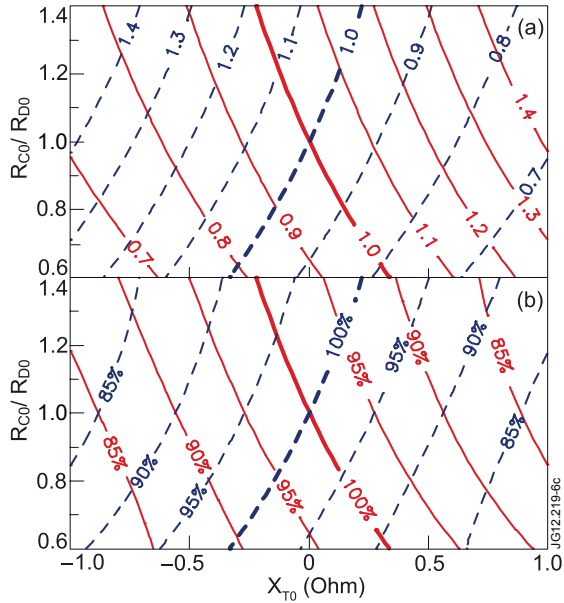

**Figure 6.** Simulations of the ECT performance between ELMs in the presence of non-zero T-junction reactance and antenna strap loading imbalance. Two sets of contour plots correspond to the matching options $CT$=0 (solid lines) and $CT$=1 (dashed lines): (a) the ratio of the maximum voltages in the conjugated lines $\left|V_{CTL}^{max}\right|/\left|V_{DTL}^{max}\right|$ and (b) the system utilisation factor. See section 3.3 for the definitions and simulation parameters.

## 4. Implementation and commissioning of the ECT system at JET

*4.1 The main features of the ECT design at JET*

The principles of the ECT matching were first assessed in 2004 during low-power (<1 MW) tests [30] on a pair of adjacent straps of one A2 ICRH antenna at JET. Despite unfavourable loading configuration (mutually coupled straps of different design), high ELM-tolerance of the prototype was successfully demonstrated. The encouraging results have led to a full-scale project involving installation of the ECT system on two A2 antennas 'C' and 'D'. Together with the ILA antenna relying on the in-vessel conjugate-T scheme [10] and the 3dB couplers energising the A2 antennas 'A' and 'B' [22], the ECT system allows us to explore and compare a wide range of advanced impedance matching techniques in the same experimental environment [11].

The implementation of the ECT project at JET faced a number challenges including the requirement of seamless integration into the existing RF plant, retaining the operational capabilities of the JET ICRH system and constraints due to relatively low priorities within the JET program. Prior to the ECT installation the straps of antennas 'C' and 'D' were individually energized by the 2MW RF amplifiers matched by conventional stub-trombone tuners located in the generator hall [13]. The transmission lines in the torus hall were also equipped with the SLiding IMPedances (SLIMPs) [46] - the disused trombone-like elements fitted during an earlier RF project at JET. The four amplifiers feeding the four straps of each A2 antenna are driven by a common RF source while the adjustable phasing of the amplifier input signals ensures flexible control of the antenna radiation spectrum. The maximum voltages in the A2 antennas cannot exceed ~30-33kV due to electrical breakdown; this limits the power levels launched by the antenna in conditions of low coupling to plasma. The power request for each RF amplifier was typically <1MW (i.e. less then one half of the available power) except for rare cases of L-mode plasmas with small Radial Outer Gap (ROG) between the separatrix and the antenna limiters.

In order to facilitate the ECT commissioning in parallel with the JET research campaign and to minimize the risks to the experimental program, a decision was taken to keep the existing stub-trombone matching system operational and to create the ECT structure as a switchable connection between the conventional matching systems of antennas 'C' and 'D' (figure 7). This approach has offered many additional advantages such as retaining the capability of running the antennas independently (i.e. at different frequency, phasing, timing, waveform etc) and having a fallback option in case of a failure of one of the antennas. Practically, the transition between the ECT and conventional circuit configurations can be performed within a few minutes between the JET pulses by remote operation of the motorized in-line (S1) and change-over (S2) switches. One of the poles of each in-line switch S1 is integrated into the 90° elbow of the transmission line and it forms the T-junction when the switch is closed. A set of four change-over switches (not shown on figure 7) is also installed on the amplifier outputs allowing quick access to the whole circuit for the 4-port network analyser measurements.

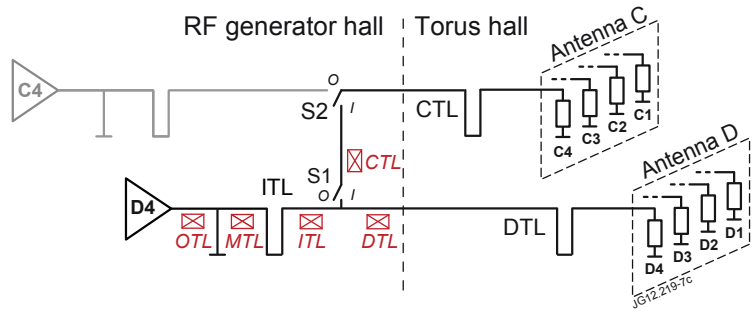

**Figure 7.** Schematic representation of one of the four identical circuits used to energize the ICRH antennas 'C' and 'D' at JET since 2008. Positions 'O' and 'I' of the in-line switch S1 and the change-over switch S2 correspond to the conventional and the Conjugate-T matching configurations respectively. The arrangement of the directional couplers used for the ECT control and diagnostic is also shown.

The ECT circuit works as a power splitter, i.e. only four amplifiers 'D1'-'D4' are used to energize the eight straps of the two antennas. The increased amplifier power demand normally doesn't introduce any additional operational restrictions because the maximum achievable antenna voltage still remains the main limiting factor in the coupling conditions typical to H-mode plasmas. In rare cases of exceptionally good coupling to L-mode plasmas the amplifier maximum power limitations could be reached first; in this case switching the system to the conventional configuration easily solves the problem. Each amplifier (e.g. 'D4') is connected to the straps occupying identical positions in different antennas (e.g. 'C4' and 'D4'). Thus the flexibility of arbitrary phasing of the straps within the antenna arrays is fully retained in the ECT configuration, while the strap phasing in the antennas 'C' and 'D' is always the same. It is worth noting, that a certain phase shift exists between the voltages on the conjugated straps (i.e. between the antennas 'C' and 'D')

$$\varphi_C - \varphi_D \approx \sqrt{\frac{2R_0}{R_{T0}}} \mp \pi \quad \text{(rad)} \tag{9}$$

where the signs correspond to the $CT=0$ and $CT=1$ matching options. This, however, has little practical significance because of the substantial spatial separation between the antennas (>3 m). For the same reason, the whole issue of mutual coupling between the conjugated straps is irrelevant to the ECT installation at JET.

Maximum effort has been made during the ECT design to reuse the existing equipment and to adapt the available diagnostic, control and protection electronics for the benefit of both matching configurations. In particular, the generator hall stubs and trombones used for conventional matching (switch positions 'O') play the role of the second stage impedance transformers in the ECT configuration (switch positions 'I'). The SLIMPs installed in the torus hall were modified and turned into proper $Z_0=30$ Ohm trombones with ~1.7m variable length such that they can be used for the ECT matching. Importantly, the real-time control algorithms for the conventional stub-trombone tuner have been shown to be adequate for the CTL and DTL trombone control in the ECT configuration (Section 5.3) which has dramatically simplified the electronics upgrade. Similarly, the new Advanced Wave Amplitude Comparison System (AWACS) for the arc detection during the ECT operations (Section 5.2) is heavily based on the conventional VSWR protection infrastructure.

*4.2 Power diagnostics and control*

The RF control and diagnostic signals for the ECT system are provided by a set of directional couplers shown in figure 7. The amplitude and phase of the forward $V^{for}$ and reflected $V^{ref}$ voltage waves are routinely monitored with 2000 samples collected for each signal per RF pulse; fast data acquisition with fixed 5μs sampling is also available for the 'OTL', 'CTL' and 'DTL' couplers. (All the RF-related experimental traces shown in the figures below were produced using the 5μs-sampled data). The couplers were individually calibrated prior to the installation and the frequency-dependent correction coefficients are applied to the raw data in order to reduce the measurement errors.

The RF power assessment using the directional couplers in the lines with high VSWR is known to suffer from very poor accuracy and in the case of the ECT installation the only reliable power measurement could be performed in the matched OTL sections on the amplifier outputs:

$$P_{OTL} = \frac{\left|V_{OTL}^{for}\right|^2 - \left|V_{OTL}^{ref}\right|^2}{2 Z_0} \tag{10}$$

The CTL and DTL couplers provide useful information about the maximum line voltages:

$$\left|V_{CTL}^{max}\right| = \left|V_{CTL}^{for}\right| + \left|V_{CTL}^{ref}\right| \quad \text{and} \quad \left|V_{DTL}^{max}\right| = \left|V_{DTL}^{for}\right| + \left|V_{DTL}^{ref}\right| \tag{11}$$

however, no accurate estimate of the $P_{CTL}$ or $P_{DTL}$ could be made using the available diagnostics and the exact power balance between the conjugated lines remains unknown. This presents certain challenges for the ECT characterisation: in general, the power balance could be affected by several independent factors such as the presence of the imaginary part in the T-junction reference impedance, a difference in loading of the two conjugated straps and a deviation of the matching trombones from the target lengths. At the same time, the maximum voltage measurements do not allow to discriminate between the origins of the asymmetry:

$$\frac{P_{CTL}}{P_{DTL}} \frac{R_D}{R_C} = \frac{\left|V_{CTL}^{max}\right|^2}{\left|V_{DTL}^{max}\right|^2} \tag{12}$$

In this situation, the individual description of coupling properties of the two straps requires the knowledge of the running value of the complex impedance $Z_T$ at the T-junction (not to be mixed up with the fixed reference value $Z_{T0}$). The measurement of the real part of this impedance is fairly straightforward:

$$R_T = \text{Re}(Z_T) \approx \frac{2 Z_0^2 P_{OTL}}{\left|V_{ITL}^{max}\right|^2} \tag{13}$$

where $\left|V_{ITL}^{max}\right|$ is the maximum voltage amplitude in the second-stage Impedance Transformer Line (ITL)

$$\left|V_{ITL}^{max}\right| = \left|V_{ITL}^{for}\right| + \left|V_{ITL}^{ref}\right| \tag{14}$$

An accurate evaluation of the imaginary part of the $Z_T$ is far more complicated: not only do the phases of the ITL voltages have to be measured but also a precise RF model of the installed T-junction assembly has to be available and incorporated into the data-processing software. Practically, these efforts for individual characterisation of the conjugated straps were not found worthwhile.

Indeed, it could be shown that the calculation of the crucially important parameter of the total power coupled to plasma by both conjugated straps doesn't require the knowledge of the power balance between the straps and their individual coupling properties; besides the specifics of the ECT design (conjugation of mutually uncoupled antenna straps with similar design and position within separate antennas) justify the assumption that both the radiation resistances and Ohmic losses in the two straps are fairly similar. In these conditions a generalized loading resistance could be defined as

$$R \approx R_C \approx R_D \approx \frac{2 Z_0^2 P_{OTL}}{\left|V_{CTL}^{max}\right|^2 + \left|V_{DTL}^{max}\right|^2} \tag{15}$$

Using this formula, the generalized resistances $R^{loss} \approx R_C^{loss} \approx R_D^{loss} \approx 0.4 - 0.7$ Ohm representing Ohmic losses in the antenna straps and transmission lines have been evaluated during the ECT high-voltage commissioning in absence of plasma at all the operational frequencies and for all four conjugated pairs. This allowed implementing the following algorithm for calculation and real-time control of the total power coupled to plasma during the ECT operations:

$$P_{ECT} = \sum_{j=1}^{4} P_{OTLj}\left(1 - \frac{R_j^{loss}}{R_j}\right) = \sum_{j=1}^{4}\left(P_{OTLj} - R_j^{loss}\frac{\left|V_{CTLj}^{max}\right|^2 + \left|V_{DTLj}^{max}\right|^2}{2 Z_0^2}\right) \tag{16}$$

where the indices $j = 1...4$ denote the lines corresponding to the four amplifiers 'D1-D4' involved in the ECT installation. For clarity, some additional small (~1.5%) implemented corrections accounting for the power losses in the second-stage impedance transformer are omitted in the considerations above.

In practice, both the open and closed $P_{ECT}$ feedback loops are used, depending on the experimental requirements. The characteristic 2 ms time-step of the existing real-time power control software is comparable with the ELM duration; in this situation, it was found essential to smooth the feedback signals in order to prevent a lagged compensation of transient $P_{ECT}$ increase during ELMs due to the momentary improvement of antenna coupling to plasma (see Section 5.1).

*4.3 Operational frequency bands*

The JET RF amplifiers have an operational frequency range of 23-57 MHz which is divided into eight 4 MHz-wide sub-bands and one (39-41 MHz) dead-band [13]. Distinctly different settings of the amplifier inter-stage and end-stage tuning elements are used within each sub-band and operations close to the sub-band edges are usually avoided. Another important practical restriction on the frequency choice is related to matching the antenna and amplifier impedances. Limited variable length (~1.5m) of the trombones used in the matching circuits doesn't ensure the required impedance transformation at arbitrary frequency. In the presence of long transmission lines, adequate matching using the conventional stub-trombone circuits could be achieved at JET only within a series of ~1.8 MHz-wide windows covering the whole operational band. Such windows have to overlap for all the amplifiers feeding the straps of the same antenna and driven by one master oscillator; this requires accurate equalisation of the transmission line lengths during the installation accounting for differences in the antenna strap equivalent electrical lengths.

Implementation of the switchable ECT system at JET faced even bigger challenges for the multi-frequency operations. Firstly, the fixed electrical lengths between the T-junctions and the antenna strap short-circuits had to be equalized for all the eight lines involved (figure 7). This is essential for having the four conjugate-T circuits tuned to identical T-junction impedances $Z_{T0}$ in the same narrow frequency windows allowed by the CTL and DTL trombone length variation. Secondly, the line lengths between the T-junctions and the stubs had to be appropriately selected and equalized in order to define the frequency windows where the second-stage impedance transformation $Z_{T0} \to Z_0$ is feasible given the limited variation of the ITL trombone lengths. The intersection of the 'allowed frequency' windows for the conjugate-T and the second-stage impedance transformer circuits defines the frequencies where the whole ECT system could be matched.

Optimisation of the positions of the 'allowed frequency' windows was the main focus of the ECT transmission line design. Dedicated network analyser measurements were performed to identify the electrical lengths of the key components of the system prior to the installation. Modifications of existing transmission line layout were kept to a minimum in order to retain the capabilities of independent conventional stub-trombone matching of the antennas 'C' and 'D'. The available options for positioning of the switches in the lines were severely restricted by accessibility issues. It was possible to install the in-line switches S1 (i.e. the T-junctions in the ECT configuration) in the same positions along the lines ~13 m from the stubs. At the same time, the change-over switches S2 had to be fitted at different positions along the lines; in this situation the line lengths (~87 m) between the T-junctions and the short circuits of the antennas 'C' and 'D' were equalized by adjusting the lengths of the fixed sections of the lines between the switches S1 and S2. Figure 8 demonstrates the results of the network analyser measurements of the equivalent line lengths for the extreme positions of the CTL and DTL trombones performed after the installation. The peculiar length dependence on frequency seen on figure 8 is explained by the resonant nature of the antenna circuit and by the presence of multiple reflective perturbations along the transmission lines (elbows, ceramics, trombones, switches etc). It can be seen from the plots that the resulting lengths are quite similar for all the lines which indicates that adequate overlapping of the 'allowed frequency' windows could be achieved for the four conjugate-T circuits.

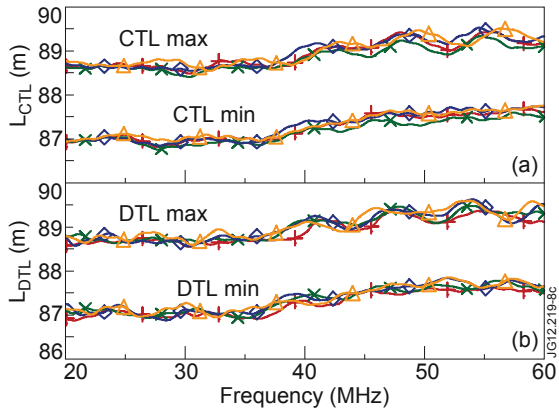

**Figure 8.** The frequency dependence of the total equivalent length of the lines between the T-junctions and the antenna strap short-circuits measured by the network analyser for extreme settings ('0 m' and '1.7 m') of the CTL (a) and DTL (b) trombones. The lengths of the lines belonging to the four RF amplifiers involved in the ECT installation are indicated by corresponding symbols.

The resulting distribution of the ECT matching frequency windows across the 23-57 MHz band is illustrated on figure 9 for the practically relevant case of antenna strap loading $R_{C0}=R_{D0}=1.5\,\Omega$ and the T-junction reference impedance $Z_{T0} = 4 + i \cdot 0\,\Omega$. The horizontal bars represent the frequencies where the four conjugate-T circuits (top), the four second-stage impedance transformers (middle) and the complete ECT system (bottom) could be set to achieve the required matching using the available range of variation of the installed trombones and stubs. The position of the bands is calculated for the alternative CT and IT matching options (indicated to the right of the plot) on the basis of network analyser measurements of the line lengths. The amplifier sub-bands and the commissioned ECT operational frequencies are also shown for reference.

Due to the significant CTL and DTL total lengths (~87 m) as compared with their range of variation (1.7m), the 'allowed frequencies' for the conjugate-T circuit evenly cover the 23-57 MHz band in a series of narrow windows; importantly both the $CT=0$ and $CT=1$ matching options are accessible at the same frequencies which offers useful opportunity for optimisation of the ECT performance as discussed in Section 3.3. The positions of the windows are not very sensitive to the antenna strap resistance or the $Z_{T0}$, thus the conjugate-T part of the ECT installation on its own doesn't limit the choice of operational frequencies. The situation is quite different for the second-stage impedance transformer where the 'allowed frequency' windows are wide and sparse. This is explained by the fact that the total ITL length is relatively small (~13 m); the variable length of the ITL trombone is 1.5 m and the stub length can be changed in the range of 0.13-3.0 m. Thus substantial gaps exist in the resulting distribution of the matching windows of the ECT system as a whole. Exploiting the second matching option $IT=1$ of the impedance transformer (see Section 3.1) allows an expansion of the range of 'allowed frequencies' and enables operation in the ~47 MHz sub-band. However, the present ECT installation remains incompatible with the ELM-tolerant operations at low frequencies <28 MHz and at the frequencies in the ~37 MHz sub-band. In accordance with the requirements of the JET experimental program, four representative frequencies were selected for the high-power commissioning of the ECT system: 32.5 MHz, 42.5 MHz, 46.0 MHz and 51.0 MHz.

It should be noted that the ECT frequency restrictions mentioned above stem primarily from the difficulties of re-configuring the long-established layout of the transmission lines at JET which made optimisation of the T-junction position impracticable. A viable solution for recovering the full frequency range has been identified which involves installation of trombones with 6 m variable length in the second-stage impedance transformer; such modifications could be implemented in future if the ECT operational frequency gaps prove to be a problem. It is also worth mentioning that the frequency restrictions are not applicable to the case when the T-junction impedance is set to $Z_{T0}=Z_0$: in this situation the second-stage impedance transformation is not needed, the stub lengths could be set to the $\lambda/4$ and the value of the ITL trombone length becomes irrelevant. This mode of the ECT matching has been successfully demonstrated at JET at the frequencies of 25.6 MHz, 29.0 MHz and 37.5 MHz, however this option offers little practical benefit for the H-mode plasma operations because of insufficient ELM-tolerance (see Section 3.2).

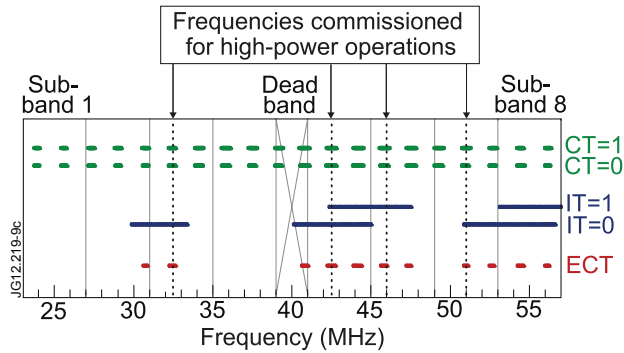

**Figure 9.** The matching frequency bands for the conjugate-T circuits (top), the second-stage impedance transformers (middle) and the complete ECT system (bottom). The amplifier operational sub-bands and the commissioned ECT frequencies are also shown for reference.

*4.4 Commissioning procedures*

The ECT system has been brought into operation in several stages which included
1. Low power pre-tuning of individual circuits using the network analyser;
2. Matching refinement, high-voltage tests and commissioning of the upgraded electronics during remotely-controlled 'asynchronous' RF pulses in the absence of plasma;
3. Plasma operations during L-mode discharges with variable antenna-plasma distances for an assessment of the real-time matching algorithms;
4. The RF power injection into ELMy H-mode plasmas for characterisation of the ECT ELM-tolerance and checking the efficiency of the new AWACS arc-detection system;
5. Maximisation of the ECT high-power capabilities by fine-tuning the circuits aiming at equalisation of the CTL and DTL voltages.
It should be noted that the last-mentioned activity still remains an essential part of the ECT optimisation during specific discharge scenarios.

The high sensitivity of the ECT performance to the T-junction reference impedance $Z_{T0}$ (Section 3.2-3.3) and the anticipated difficulties of accurate in-situ $Z_T$ measurements using the directional couplers (Section 4.2) provided strong motivation for a detailed dedicated investigation of the parameters of the second-stage impedance transformer prior to its installation. An exact replica of the transformer was assembled, and the 2-port S-matrix of the circuit was measured using the network analyser over the entire frequency band which allowed the establishment of the stub and trombone lengths required for the $Z_{T0} \to Z_0$ impedance transformation. With these settings available, further low-power tuning of the full ECT assembly involved finding the CTL and DTL trombone lengths corresponding to the OTL VSWR $S \approx 1$ in absence of plasma.

One of the important issues assessed during the ECT vacuum tests was the system vulnerability to errors of the matching element control arising from the finite accuracy of the actuators. Theoretical studies indicate that the acceptable (OTL VSWR $S<3$) tolerance of the settings near the perfect matching point is inversely proportional to the frequency and it also quickly decreases with the reduction of the antenna loading. In the in-vessel conjugate-T circuits the required tolerances of the capacitor control reach deep into the technologically challenging sub-millimetre range [37] even in the presence of plasma, let alone the vacuum case. The sensitivity of the ECT matching to the CTL and DTL trombone lengths predicted by the simulations and confirmed during the tests was found to be quite manageable. Figure 10 shows the calculated OTL VSWR dependence on the CTL and DTL lengths for a case of $f$=51 MHz, $Z_{T0} = 4 + i \cdot 0$ Ohm and the ECT resistive loads representative of (a) vacuum $R_{C0}=R_{D0}$=0.5 Ohm and (b) plasma conditions $R_{C0}=R_{D0}$=1.5 Ohm. The two points corresponding to perfect ($S$=1) matching identify the CT matching options discussed in Section 3.1 and the bold contours ($S$=3) indicate the boundaries of the acceptable length deviation from the target values. It can be seen from figure 10(a) that even in the most demanding practical case of high frequency operations in vacuum the required matching tolerances are well within the existing capabilities of the trombone length control (≤0.5 cm). Plasma loading allows much bigger deviation of the trombone lengths from the target values (figure 10(b)) which provides good margin for stable operations of the real-time matching algorithms (see Section 5.3).

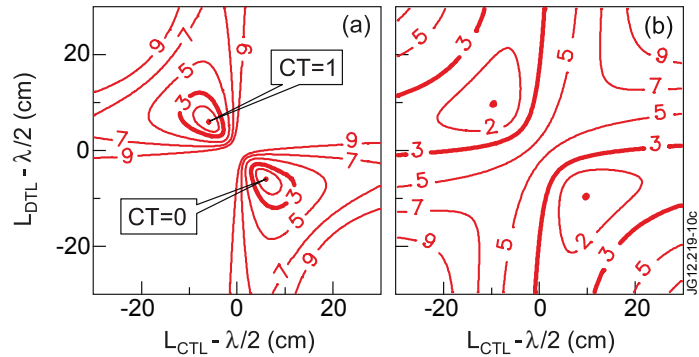

**Figure 10.** The OTL VSWR dependence on the CTL and DTL lengths for a case of $f$=51 MHz, $Z_{T0} = 4 + i \cdot 0$ Ohm and the ECT resistive loads representative of (a) vacuum $R_{C0}=R_{D0}$=0.5 Ohm and (b) plasma conditions $R_{C0}=R_{D0}$=1.5 Ohm.

The transition from the vacuum tests to plasma operations was found to be straightforward: RF power of ~1 MW was coupled to plasma without trips during the very first ECT plasma discharge with the L-H transition and ELMy H-mode (pulse #76606). The vacuum matching lengths of the CTL and DTL trombones proved to be adequate for plasma operations providing they are increased by ~10 cm to account for shortening of the equivalent length of the antenna straps in the presence of plasma. At the same time, the ECT system generally demonstrates lower sensitivity to the accuracy of the initial settings of the matching elements as compared with the conventional stub-trombone circuit. This facilitates RF operations in conditions when different plasma scenarios are used during the experimental session. Together with the capability of switching to the conventional matching scheme, this flexibility allowed the ECT commissioning to be performed as a background task during the main research program at JET. No difficulties were encountered during the changes of the antenna phasing and by now the ECT system has been used with (0,π,0,π), ±(0,π/2,π ,3π/2) and (0,π,π,0) phase shift between the antenna strap voltages.

Presently the ECT system is fully integrated into the RF plant local & remote control and it has been routinely used during the RF operations since 2009 delivering up to 4 MW into ELMy plasmas. The ECT performance demonstrated in plasma conditions, including the issues of ELM-tolerance, arc detection and real-time matching, is discussed below together with some further details of the system design.

## 5. The ECT performance during plasma operations

*5.1 ELM-tolerance*

Experiments in H-mode plasma conditions have fully confirmed that the ECT system achieves its ultimate goal of substantially reducing the magnitude of impedance perturbations on the output of the RF amplifiers during strong and fast antenna loading changes introduced by the ELMs. The maximum VSWR values measured in the amplifier Output Transmission Line (OTL) using data acquisition with 5μs sampling did not exceed $S \approx 1.5-2$ despite momentary variation of the CTL and DTL loading resistance in the range of $R \approx 1-8\,\text{Ohm}$. The antenna strap inductance changes accompanying the ELMs and manifesting themselves as rapid perturbations of the CTL and DTL equivalent lengths $\delta L$ have not been found critically detrimental to the ELM-tolerance (see discussions in Section 3.3); so far no dedicated adjustments of the original $\text{Im}(Z_{T0}) \approx 0$ impedance settings have been necessary.

A typical example of the VSWR response to an ELM in one of the ECT constituent circuits is shown on figure 11. Note that the loading resistance plotted on figure 11(b) and used in other experimental figures in this paper corresponds to the generalized definition given by formula (15) in Section 4.2 and it accounts for both the RF power coupling to plasma and the Ohmic losses in the antenna and transmission lines. The CTL and DTL equivalent length perturbation $\delta L$ shown in the figures below is averaged between the two lines. The length perturbation $\delta L$ is deduced from the phases $\varphi^{\text{for}}$ and $\varphi^{\text{ref}}$ of the forward and reflected voltage waves which are routinely measured by the CTL and DTL directional couplers (see figure 7): $\delta L = (\delta\varphi^{\text{for}} - \delta\varphi^{\text{ref}})\lambda/(4\pi)$, where $\delta\varphi^{\text{for}}$ and $\delta\varphi^{\text{ref}}$ are the phase perturbations expressed in radians and λ is the wavelength. It is of interest that the temporal behaviour of the OTL VSWR (figure 11(d)) has two peaks of equal magnitude where the first one is much sharper; this is consistent with the theoretical curve shown on figure 4 and with the fact that the VSWR goes through the local maximum twice as the loading resistance rapidly increases and slowly decreases during the ELM. Further efforts reproduced the aforementioned curve explicitly and for a broader range of loading resistances including the small values below the reference resistance $R_0$. This proved possible during the ELMy pulse #77293 when the Radial Outer Gap (ROG) between the antenna and plasma separatrix was slowly increased up to very large values $G$=14cm while the real-time control of the matching trombones was disabled. The RF data collected in these conditions during ~200 consecutive ELMs are presented on figure 12. The plots confirm that the ECT circuit demonstrates the characteristic ELM-tolerant features predicted by the theory (figure 4) and that its experimental response (figure 12(c)) is in good agreement with the results of simulations (figure 12(b)) based on realistic input parameters (figure 12(a)).

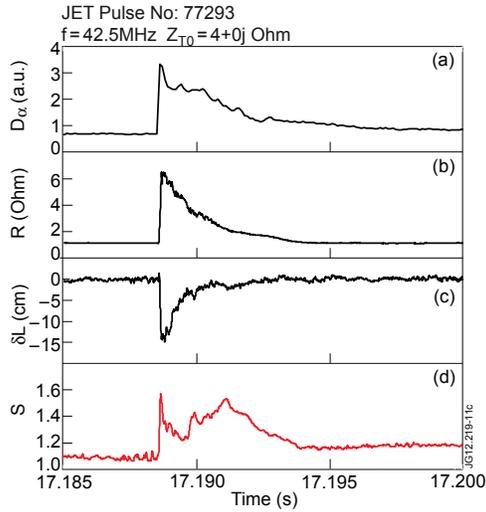
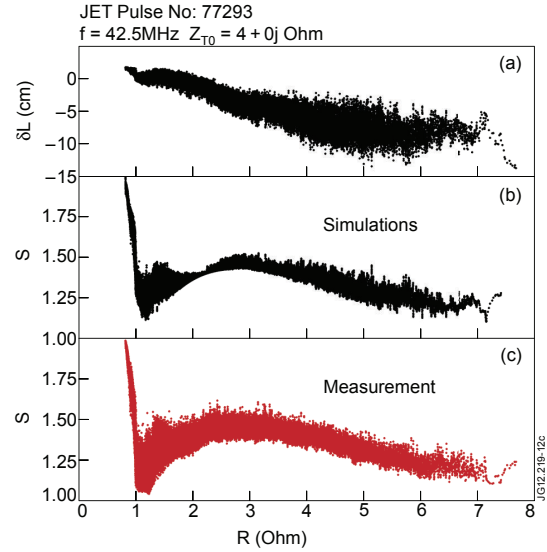

**Figure 11.** The temporal response of the VSWR in the Output Transmission Line (OTL) of the amplifier 'D4' to the antenna loading perturbation during an ELM: (a) intensity of $D_\alpha$-line emission from plasma indicating the occurrence of an ELM, (b) loading resistance, (c) the CTL and DTL equivalent length perturbation by the ELM and (d) the OTL VSWR.

**Figure 12.** The ECT matching tolerance to the antenna loading perturbations during ELMs: (a) the measured dependence of the CTL and DTL equivalent length perturbation on the loading resistance, (b) the OTL VSWR response to the loading perturbations above as predicted by the circuit simulations and (c) the measured OTL VSWR. The points correspond to the 5µs-sampled data collected during ~200 consecutive ELMs on the ECT transmission lines energized by the amplifier 'D4'.

The most important practical consequence of the enhanced ELM tolerance of the ECT system is that the VSWR in the OTL remains well below the trip threshold of the RF amplifier protection ($S$=3 at full power) during H-mode plasma operations. This ensures uninterrupted performance of the RF plant in the presence of ELMs and thus improves the reliability of the RF power control and noticeably increases the average power levels deliverable to ELMy plasmas. Figure 13 provides a vivid comparison of the OTL VSWR and the coupled power waveforms recorded during two identical ELMy plasma pulses where the RF amplifiers 'D' were running in the conventional and the ECT configuration (switch positions 'O' and 'I' respectively on figure 7). The described improvement in the operational capabilities of the RF plant has been observed at all the commissioned frequencies, at different phasing of the antenna straps and in a variety of plasma discharges with rapidly changing parameters affecting the antenna loading (e.g. 'sawtooth' activity, L-H mode transitions, Type-I and Type-III ELMs etc). Figure 14 demonstrates successful trip-free performance of the ECT system at $f$=51.0 MHz during JET high-energy pulse #78125 ($I_p$ = 4.15 MA, $P_{NBI}$ = 22 MW, $W_{DIA} \approx 11.5$ MJ) with big $\Delta W_{DIA} \approx 0.6$ MJ ELMs.

One of the peculiarities of the ECT implementation at JET is that the power coupled to plasma experiences sharp increase during the ELMs (figure 14(e)). Such behaviour arises from a rapid (<1ms) reduction of the RF power losses in the long transmission lines under the high loading conditions during the ELMs while the real-time power control software is not fast enough to adjust the applied power. These ELM-related surges of coupled power could be seen as a manifestation of higher system efficiency under high loading and they don't present any danger to the equipment; on the contrary, this phenomenon could make a noticeable contribution to the average power levels coupled to H-mode plasmas with frequent ELMs. The last observation indicates that the ECT system has certain advantage over the alternative 3dB hybrid system where a part of the RF power is diverted from the antennas and lost in the dummy loads during the ELMs thus reducing the average power levels coupled to plasma.

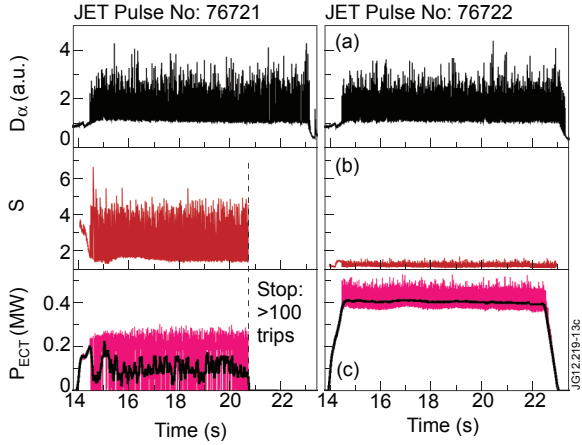
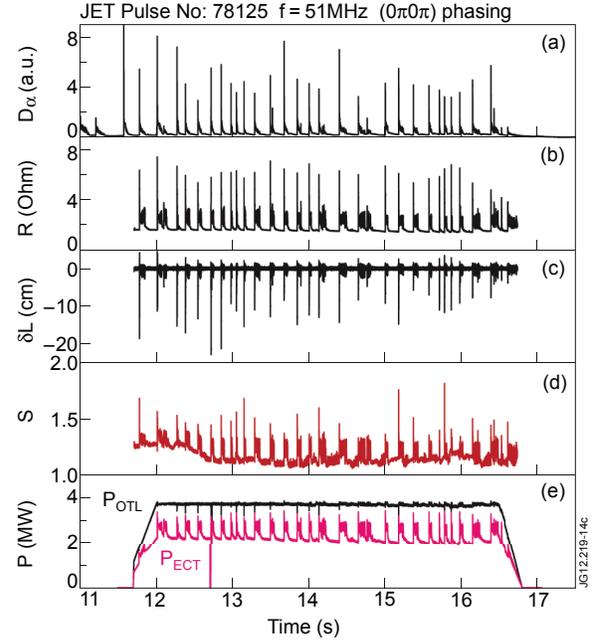

**Figure 13.** Comparative performance of the RF amplifiers 'D' in the conventional (left) and the ECT (right) matching configurations during two similar ELMy plasma discharges: (a) intensity of $D_\alpha$-line emission from plasma, (b) the OTL VSWR and (c) instantaneous and 0.1s window-averaged RF power coupled to plasma. Traces (b) and (c) correspond to the data averaged over the four constituent circuits. Operations at 42.5 MHz with $(0\pi0\pi)$ phasing of antenna straps.

**Figure 14.** The ECT behaviour during H-mode plasma with big ELMs: (a) intensity of $D_\alpha$-line emission from the plasma, (b) loading resistance, (c) the CTL and DTL equivalent length perturbation by ELMs (d) the OTL VSWR, (e) the RF power generated and coupled to the plasma by the ECT system. The traces (b), (c) and (d) represent the values averaged over the four constituent ECT circuits.

*5.2 Arc protection*

In the conventional configuration of the RF plant (switch positions 'O' on figure 7) the antenna and transmission line protection against RF arcs is traditionally provided at JET by a combination of two methods [11,13]: the applied RF power is automatically limited when the maximum voltages in the Main Transmission Lines (MTL) reach a pre-set value (typically 30-33kV) and the applied power is tripped for a short period of time when the VSWR in the amplifier Output Transmission Line (OTL) exceeds a power-dependent threshold (typically $S\geq 3$). It should be noted that the VSWR trip system has a dual purpose as it also serves as a part of the end-stage tetrode protection; the choice of the trip threshold is a compromise between the RF plant protection and uninterrupted operations in presence of variable plasma loading allowing some margins for the real-time matching system to home-in. The trip (5-20ms) and recovery (10-80ms) duration depends on the frequency of the trip triggering events and the RF pulse is terminated if more than 100 trips occur.

On the whole, this approach is retained in the ECT configuration (switch positions 'I' on figure 7), at the same time, it is substantially expanded and advanced in order to meet more challenging arc protection requirements. Firstly, the automatic power limitation system is now activated when the maximum voltages are reached in any of the three lines including the CTL, DTL and ITL. Secondly, the power trip system triggered by the high VSWR values in the OTL is complemented by the triggers generated by the so-called Advanced Wave Amplitude Comparison System (AWACS) [31].

The necessity of development of the AWACS is explained by an inadequate sensitivity of the traditional high-VSWR detection method to the occurrence of arcs in the ECT circuit. Extensive computer simulations have revealed that the VSWR values in the OTL could remain quite low (below the trip thresholds) during certain realistic arcing scenarios. It should be noted, that these scenarios go far beyond the peculiar 'voltage node' arcing [47] which is notoriously difficult to detect [48]. Even the most common high voltage ('voltage anti-node') arcs in one of the conjugated lines or the antenna straps could evade detection by the VSWR technique during the ECT operations [31]; such cases include arcing in the presence of circuit asymmetries due to $R_{C0} \neq R_{D0}$ or $\text{Im}(Z_{T0}) \neq 0$, high loading resistance $R_0$ matching or trombone deviations from the target positions.

A further serious challenge specific to the ELM-tolerant systems is related to detection of arcs during the ELMs. This problem has only started to emerge recently [11,22] after the introduction of the RF systems which do not cause the RF power interruption during the development of large ELMs. The observations indicate that the electrical strength of the ICRH antennas could be adversely affected by the strong particle and radiation fluxes ejected from plasma during the ELMs which increase the probability of arcing despite reduction of the antenna voltages under the ELM-related high loading. Figure 15 provides a characteristic example of the antenna breakdown recorded during the ECT operations where an arc takes place with a brief delay (~65μs) after the beginning of an ELM; the other typical cases include arcing just at the ELM onset.

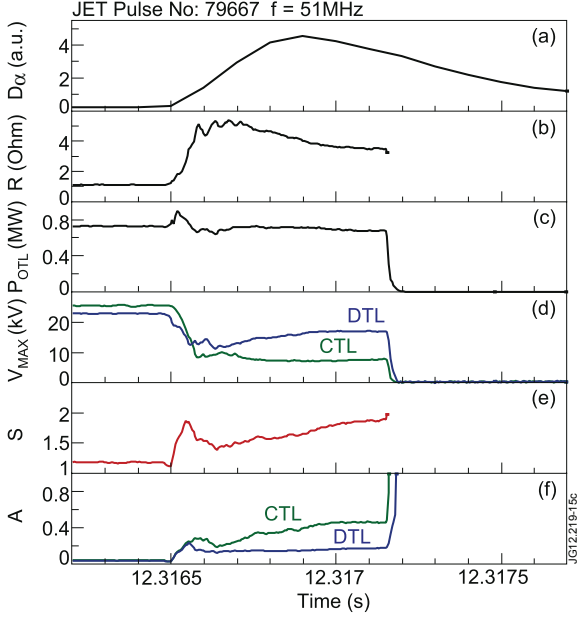

**Figure 15.** An example of arcing in the strap 2 of antenna 'C' taking place ~65μs after the beginning of an ELM and triggering the protection power trip: (a) intensity of $D_\alpha$-line emission from the plasma, (b) loading resistance, (c) output power of the RF amplifier 'D2', (d) the CTL and DTL maximum voltages, (e) the VSWR in the OTL of the amplifier 'D2' and (f) the CTL and DTL AWACS arc detection signals.

The above-mentioned shortcomings of the traditional high-VSWR arc detection method are compounded by its incapacity to discriminate the ELM-induced arcs. The results of simple simulation shown on figure 16(d) indicate that a fairly common arc taking place during an ELM at the maximum voltage location in one of the conjugated lines will not cause a sufficiently high ($S>3$) VSWR increase to trigger the protection power trip. In these simulations the arc was represented as a low-inductance (20nH) short-circuit and a simple model of an ELM with purely resistive and symmetrical loading of both conjugated lines was used; the rest of the simulation parameters were as follows: $f$=42.5MHz, $R_{C0}$=$R_{D0}$=1 Ohm, $Z_{T0} = 4 + i \cdot 0$ Ohm and the OTL forward power was fixed at 2 MW. In order to overcome the described difficulties an advanced version of the trip system has been proposed for the ECT project where the OTL high-VSWR trigger signals are complemented by two signals $A_{CTL}$ and $A_{DTL}$ produced as the ratios of the OTL reflected voltage amplitude and the forward voltage amplitude in the CTL and DTL respectively:

$$A_{CTL} = \frac{\left|V_{OTL}^{ref}\right|}{\left|V_{CTL}^{for}\right|} \quad \text{and} \quad A_{DTL} = \frac{\left|V_{OTL}^{ref}\right|}{\left|V_{DTL}^{for}\right|} \tag{17}$$

The simulations (figure 16) demonstrate that such AWACS signals are much more sensitive to the presence of arcs in the CTL or DTL as compared with the OTL VSWR: indeed, not only does the nominator of the ratio increase during the arc (figure 16(a)) but also the denominator of the ratio in the affected line experiences a noticeable decrease (figure 16(b)) while it is not the case for the second line of the conjugated pair (figure 16(c)). Such behaviour of the CTL and DTL forward voltages is explained by a strong increase of the parallel impedance presented to the T-junction by the arcing line. As a result, the margin between the normal and arc-related levels of the AWACS signals (figure 16(e) and (f)) is much wider than it is for the OTL VSWR (figure 16(d)) both during ELMs and in other operational conditions [31]. Setting the AWACS trip thresholds within this margin, e.g. $A_{CTL}^{trip} = A_{DTL}^{trip} \approx 0.8$ as adopted in the ECT design, allows reliable arc detection without triggering spurious trips related to ELMs or other circuit perturbations not related to the electrical breakdown. Also important is that, unlike the VSWR method, the AWACS is capable of identifying which of the conjugated branches, the CTL or the DTL, is affected by arcing.

At the same time, it was found useful to keep the traditional high-VSWR triggering in the trip system because of its higher sensitivity to the arcs in the ITL and to simultaneous arcs in both the CTL and DTL branches; finally the VSWR trips remain the main protection of the amplifier end-stage tubes against an accidental impedance mismatch due to wrong setting of the trombone lengths.

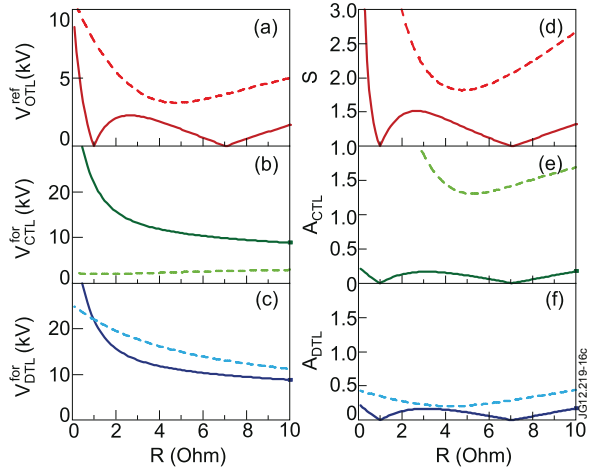

**Figure 16.** Simulations of the ECT behaviour during the ELM-related variation of the loading resistance in the absence and in the presence of arcing (solid and dashed curves respectively) at the maximum voltage location in the CTL: (a) reflected voltage amplitude in the OTL, (b) the CTL forward voltage amplitude, (c) the DTL forward voltage amplitude, (d) the VSWR in the OTL, (e) the CTL AWACS arc detection signal and (f) the DTL AWACS arc detection signal.

An experimental example of arc detection during the development of ELMs is given in figure 17. The plots demonstrate the expected low sensitivity of the VSWR in the OTL to the occurrence of an arc; in contrast, the $A_{DTL}$ ratio shows a sharp increase and triggers the protection power trip. The $A_{CTL}$ ratio remains relatively unaffected which allows positive attribution of the arcing event to the antenna 'D'. In general, all the experience of the ECT operations so far confirms the high efficiency of the AWACS arc detection both between and during ELMs and the instances of false triggering due to 'benign' loading perturbations are extremely rare.

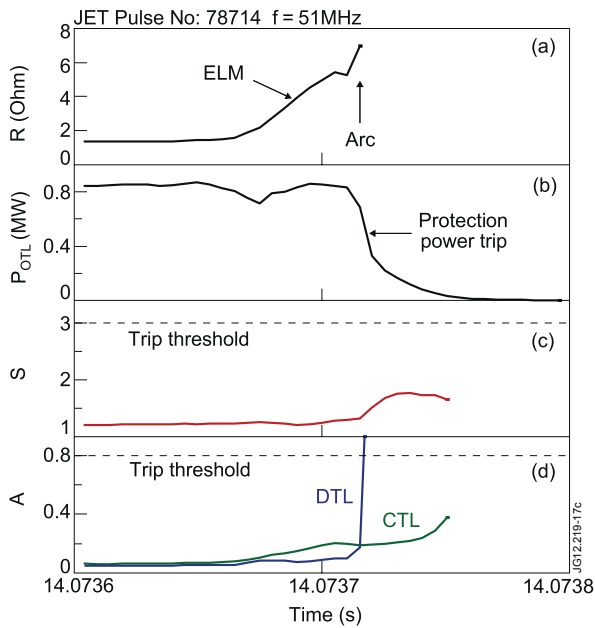

**Figure 17.** An example of operations of the AWACS protection during the ELM-induced arcing in the strap 1 of antenna 'D': (a) loading resistance, (b) output power of the RF amplifier 'D1', (c) the VSWR in the OTL of the amplifier 'D1', (f) the CTL and DTL AWACS arc detection signals.

Practical advantages of the AWACS implementation, its simplicity and straightforward integration into the existing JET RF plant controls are also worth mentioning. The AWACS hardware is entirely based on the same electronic modules which are used for the high-VSWR triggering with the only difference that some of the input signals are changed ($V_{\text{CTL}}^{\text{for}}$ or $V_{\text{DTL}}^{\text{for}}$ is used instead of $V_{\text{OTL}}^{\text{for}}$) and the trip thresholds adjusted. This minimizes the development work and allows a compact and versatile design of the protection system.

*5.3 Real-time matching algorithms*

Due to the intrinsically high load-tolerance of the ECT system, the availability of the real-time matching control is generally not as critical as in the case of conventional matching schemes; at the same time, it remains a useful feature which facilitates transition to new experimental scenarios and helps to retain the ELM-tolerance during slow evolution of plasma parameters. In order to simplify the ECT integration into the long-established and complex infrastructure of the RF plant, maximum effort has been made to exploit the existing algorithms and electronics for the ECT control.

In the conventional matching configuration (switch positions 'O' on figure 7) the lengths of the CTL and DTL trombones are fixed and the lengths of the ITL trombone and the stub are controlled automatically in accordance with the so-called 'frequency error $E_F$' and 'stub error $E_S$' signals respectively. (The term 'frequency error' is historical: in the past the error signal $E_F$ was also used for the real-time frequency adjustments; this option is now disabled). The error signals $E_F$ and $E_S$ are derived from the amplitudes and the phases of the MTL forward $V_{\text{MTL}}^{\text{for}}$ and the OTL reflected $V_{\text{OTL}}^{\text{ref}}$ voltages measured by the directional couplers installed equidistantly from the stub junction (see figure 7):

$$E_F[\text{Volt}] = -20 \, \text{Re}\left(\frac{V_{\text{OTL}}^{\text{ref}}}{V_{\text{MTL}}^{\text{for}}}\right) = -20 \frac{|V_{\text{OTL}}^{\text{ref}}|}{|V_{\text{MTL}}^{\text{for}}|} \cos\left(\varphi_{\text{OTL}}^{\text{ref}} - \varphi_{\text{MTL}}^{\text{for}}\right) \qquad (18)$$

$$E_S[\text{Volt}] = -20 \, \text{Im}\left(\frac{V_{\text{OTL}}^{\text{ref}}}{V_{\text{MTL}}^{\text{for}}}\right) = -20 \frac{|V_{\text{OTL}}^{\text{ref}}|}{|V_{\text{MTL}}^{\text{for}}|} \sin\left(\varphi_{\text{OTL}}^{\text{ref}} - \varphi_{\text{MTL}}^{\text{for}}\right) \qquad (19)$$

Close to the matching conditions $V_{\text{OTL}}^{\text{ref}} = 0$, the amplitudes of the error signals indicate (although not linearly) the magnitude of deviation of the stub and trombone lengths from the target values and their signs define the directions of the length change required for achieving the impedance matching, e.g. positive $E_S$ signal means that the stub has to be shorten etc [13]. The rate of the element length variation is normally set to the maximum and it doesn't change except for a narrow range of error signals close to zero where the rate is reduced before the elements are stopped. The algorithm and the associated electronics have been successfully used for the real-time matching control at JET since the implementation of the stub-trombone scheme.

At first glance, the control technique described above doesn't look suitable for the ECT circuit configuration (switch positions 'I' on figure 7). Indeed, in this case it is the lengths of the CTL and DTL trombones which have to be adjusted in real time while the stub and the ITL trombone lengths have to remain fixed: this requirement is essential for keeping the T-junction reference impedance constant and optimal for the ELM-tolerance during the automatic load tracking. In-depth analysis has shown, however, that the error signals $E_F$ and $E_S$ produced for the conventional matching scheme and the existing control infrastructure could be easily adapted to the new circumstances [31]. This conclusion is based on the fact that in the ECT circuit the 'frequency' and 'stub' error signals are proportional to deviation of respectively the imaginary and real parts of the T-junction impedance from the reference value $Z_{T0}$, hence the CTL and DTL lengths, which define the T-junction impedance, could be controlled by the $E_F$ and $E_S$ signals. In broader terms, the discussed algorithm could be considered largely universal regardless of the specific implementation or the physical location of the matching elements providing they are capable of quasi-independent variation of the real and imaginary parts of the impedance.

As applied to the ECT configuration at JET, the 'frequency' and 'stub error signals are found entirely adequate for the real-time control providing their signs and allocation to the CTL and DTL trombones are set consistent with the chosen CT and IT options; furthermore, two options exist for the coordinated CTL and DTL trombone length response to the error signals which ensure automatic matching. These options, referred to as the Tracking Modes $TM$=0 and $TM$=1, are explained in figure 18 where typical dependences of the error signal $E_F$ and $E_S$ signs on the CTL and DTL lengths are shown and the directions of the trombone length changes are indicated by the arrows. The horizontal and vertical hatching on figure 18 denotes the operational space with the negative signs of the error signals $E_F$ and $E_S$ respectively while the bold lines indicate the boundary 'zero error' conditions. The 'clockwise' and 'counter-clockwise' sets of arrows schematically illustrate the coordinated directions of the CTL and DTL length changes for the two implemented automatic Tracking Modes $TM$=0 and $TM$=1. Note, that the CTL and DTL trombones have an identical rate of length variation (~10cm/sec), therefore the arrows have the angles of $\pm 45°\pm 180°$ with respect to the axes. The OTL VSWR distribution is also presented on the figure as thin contours. The simulations correspond to the case of $f$=42.5 MHz, $Z_{T0} = 4+i\cdot 0$ Ohm, $R_{C0}$=$R_{D0}$=1.5 Ohm and the impedance transformer options $IT$=0 (figure 18(a)) and $IT$=1 (figure 18(b)).

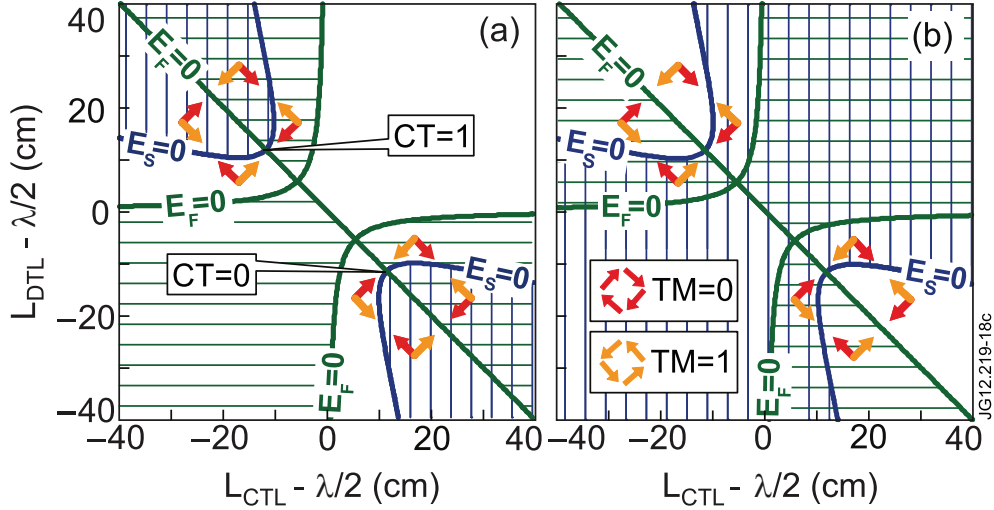

**Figure 18.** The adopted logic of automatic control of the CTL and DTL trombone lengths with respect to the signs of the conventional error signals $E_F$ and $E_S$ in the vicinity of two CT matching solutions: (a) the impedance transformer option $IT$=0 and (b) the impedance transformer option $IT$=1. See section 5.3 for detailed explanation of the plot features.

Practically, the control algorithm shown in figure 18 is implemented using a simple commutation block added to the existing electronics and allocating the error signals $E_F$ and $E_S$ to the CTL and DTL trombones according to the following logic:

$$E_{CTL} = (-1)^{IT+CT+TM+1} E_S \qquad E_{CTL} = (-1)^{IT+1} E_F$$
$$\text{if } TM=0 \text{ or } \qquad\qquad \text{if } TM=1 \qquad (20)$$
$$E_{DTL} = (-1)^{IT+1} E_F \qquad E_{DTL} = (-1)^{IT+CT+TM+1} E_S$$

where $IT$, $CT$ and $TM$ are the binary values of the ECT control options defined in this paper and the $E_{CTL}$ and $E_{DTL}$ are the error signals used to control the lengths of CTL and DTL trombones respectively; the positive or negative signs of the error signals trigger the reduction or increase of the trombone lengths correspondingly.

Figure 19 illustrates the anticipated CTL and DTL length response to the error signals $E_{CTL}$ and $E_{DTL}$ for different initial length deviations from the matching solution $CT$=1 and corresponding to the tracking modes $TM$=0 (figure 19(a)) and $TM$=1 (figure 19(b)). The simulations were performed for the impedance transformer option $IT$=0 in conditions of $f$=42.5 MHz, $Z_{T0} = 4+i\cdot 0$ Ohm and $R_{C0}$=$R_{D0}$=1.5 Ohm. The dashed lines represent two examples of the 'matching trajectories' starting from different 'guess' positions indicated by the triangle symbols. The horizontal and vertical hatching denotes the operational space with the negative signs of the error signals $E_{CTL}$ and $E_{DTL}$ used to control the lengths of the CTL and DTL trombones respectively while the bold lines indicate the corresponding 'zero error' conditions. The OTL VSWR distribution is also presented as thin contours. The CTL and DTL lengths plotted along the axes are expressed as the deviations from the matching values. It could be understood from the plots that the algorithms ensure automatic matching for any combination of the error signal signs in a wide range of initial 'guess' settings around the matching point. The exact boundaries of the region of stable automatic matching depend on a number of operational parameters and experimental conditions and their definition is not straightforward; in reality, however, these boundaries are largely overridden by the limitations imposed by the protection system and related to the maximum allowed VSWR in the amplifier Output Transmission Line (OTL). In the case of symmetrical loading of the ECT circuit $R_{C0}=R_{D0}$ and purely resistive T-junction impedance $Im(Z_{T0})=0$ no particular tracking mode or CT option have an advantage over the other in terms of reliability of the matching algorithms. In more complicated circumstances of an unbalanced ECT circuit the error signal patterns shown in figure 18 and figure 19 loose their symmetry and certain TM and CT combinations become preferable [31]. In order to enhance the capabilities of the ECT system, remote selection of the TM, CT and IT options is provided to the RF plant operator who could optimize the control mode for a particular experimental scenario. It should be noted here, that in practice the ECT real-time matching was found quite robust and not very sensitive to the choice of the TM or CT options; so far the *TM*=0 and *CT*=1 combination have been used in most cases and with little evidence of 'runaway' behaviour.

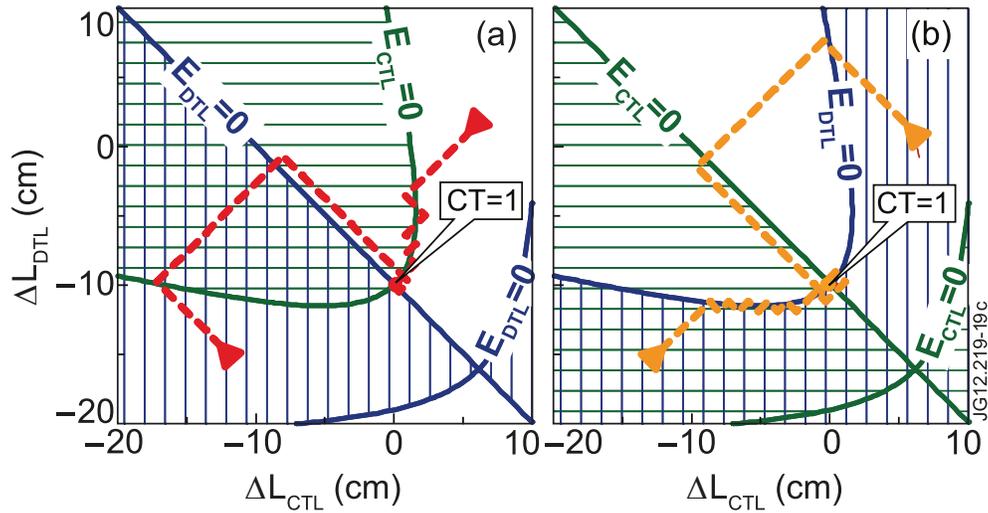

**Figure 19.** Simulations of the ECT automatic matching due the CTL and DTL length adjustments driven by the error signals $E_{CTL}$ and $E_{DTL}$: (a) Tracking Mode *TM*=0 and (b) Tracking Mode *TM*=1. See section 5.3 for detailed explanation of the plot features.

Figure 20 demonstrates an experimental example where the ECT real-time control performs an automatic 'homing-in' with subsequent tracking of variable antenna load during an L-mode plasma discharge with the ROG (Radial Outer Gap) sweep. Automatic matching in ELMy plasma conditions has been proven equally reliable; at the same time, it was found useful to implement an additional low-pass (~10Hz) filtering of the error signals in order to smooth the strong and fast ELM-related perturbations potentially stressful to the trombone drive mechanism.

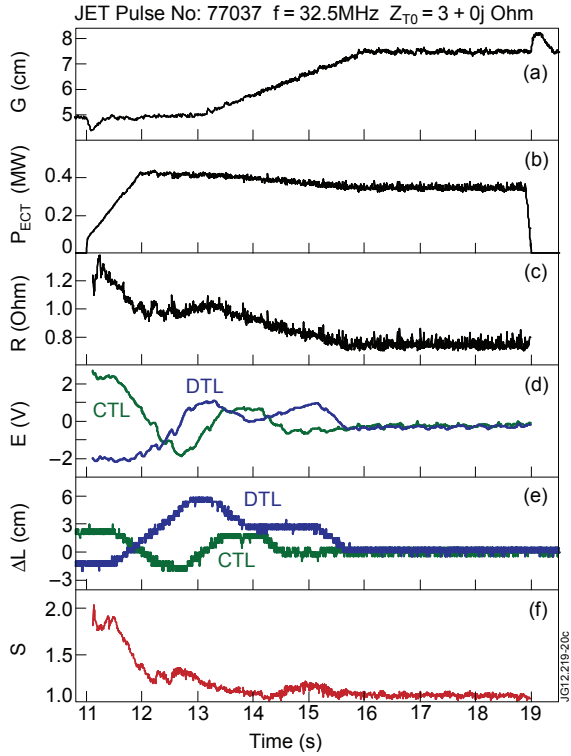

**Figure 20.** Operations of the ECT automatic matching system for the amplifier 'D2' ($IT$=0, $CT$=1, $TM$=0): (a) Radial Outer Gap between the separatrix and the antenna limiters, (b) coupled power, (c) loading resistance, (d) the CTL and DTL length error signals $E_{CTL}$ and $E_{DTL}$, (e) the deviation of the CTL and DTL trombone lengths from the matching values and (f) the OTL VSWR.

### 5.4. High power operations

The implementation of the ECT system at JET has substantially improved the RF plant performance during ELMy plasmas. Depending on the discharge scenario, the trip-free time-average power levels coupled to ELMy H-mode plasma by the ECT system alone have reached 4 MW (figure 21). Up to 7 MW total RF power has been coupled to ELMy plasma by all four A2 JET ICRH antennas using the ECT (antennas 'C' and 'D') and the 3dB coupler system (antennas 'A' and 'B') installed earlier [11,22]. Over 8 MW was injected in 2009 by all the RF systems available at the time including the ILA (figure 22). The main factor limiting further power increase remains the A2 antenna electrical breakdown at voltages exceeding 30-33 kV. The ECT has been found to be slightly outperforming the 3dB system in identical experimental conditions (e.g. see figure 22); this is explained by the peculiarities of the ECT and 3dB system's reaction to ELMs leading to opposite momentary changes of the power coupled to the plasma and by higher flexibility of the ECT circuit in equalizing the voltages on the paired antenna straps.

Although controllable in principle, the voltage imbalance in the conjugated lines has proven so far to be the most challenging practical issue for optimisation of the ECT operations at high power levels. Indeed, as soon as the maximum voltage in one strap approaches the ~30 kV breakdown levels, the ECT power becomes limited by the protection system (or by arcing) while the second strap may still be capable of safely delivering more power. Thus equalisation of the maximum CTL and DTL voltages is an important pre-requisite for utilising the full power capability of the antenna pair. In theory, the perfectly matched ECT circuit with equal CTL and DTL loading resistances $R_{C0} = R_{D0}$ and zero imaginary part of the T-junction impedance $\text{Im}(Z_{T0}) = 0$ has equal maximum voltages on the antenna straps 'C' and 'D' (figure 23) while any deviation from the above conditions creates certain voltage imbalance. As it was discussed in Section 3.3, the voltage imbalance due to the loading asymmetry $R_{C0} \neq R_{D0}$ could be compensated by the appropriate $\text{Im}(Z_{T0}) \neq 0$ adjustments; this, however, doesn't solve all the problems. It was found in practice that the accuracy of the real-time CTL and DTL trombone length control, which is quite adequate for the impedance matching, is often insufficient for ensuring an acceptable voltage balance. The simulations (figure 23) show that in some circumstances even a relatively insignificant residual impedance mismatch with the OTL VSWR values of $S$~1.2 could be accompanied by a ~20% voltage imbalance and much stronger imbalance could take place within the $S$~3 boundaries acceptable for the amplifier VSWR protection. An experimental example of the line voltage evolution during high-power ECT operations under the conditions of variable ROG is shown in figure 21(d). In this case the initial value of the VSWR in the OTL was fairly small (figure 21(e)) and didn't require adjustments of the CTL and DTL trombone lengths. At the same time quite noticeable voltage imbalance was recorded at the beginning of the pulse with the DTL maximum voltage reaching the protection limit; this imbalance has gradually disappeared by the end of the pulse consistent with the OTL VSWR reduction due to the ROG-related plasma loading changes.

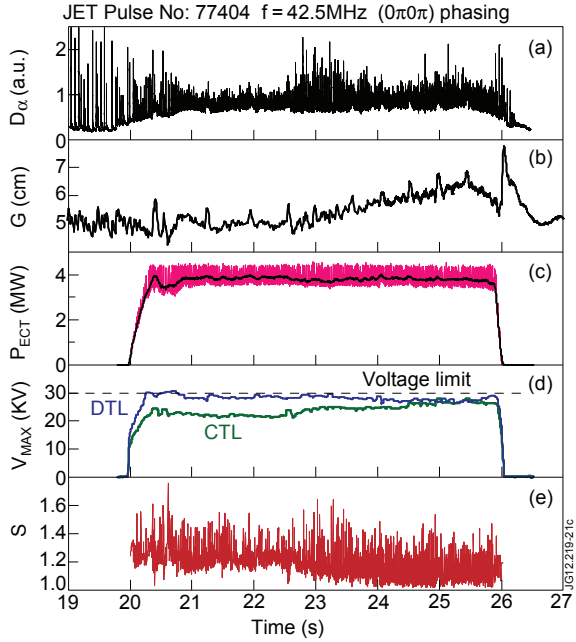
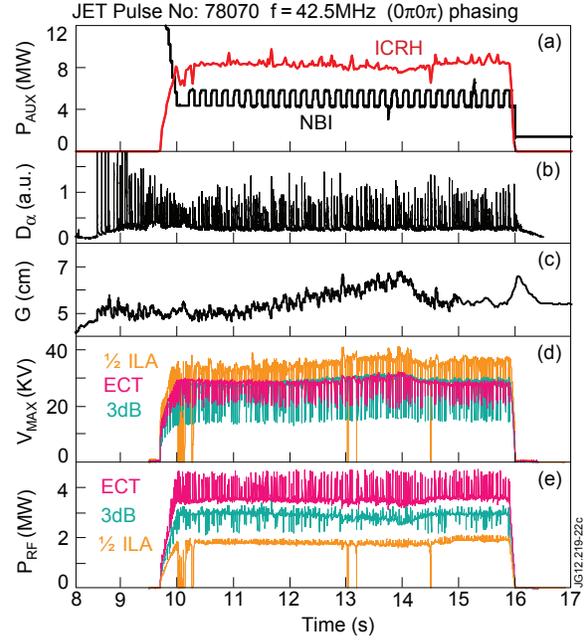

**Figure 21.** High power operations of the ECT system and the voltage imbalance in the conjugated lines: (a) intensity of $D_\alpha$-line emission from plasma, (b) Radial Outer Gap between the separatrix and the antenna limiters, (c) instantaneous and 0.1s window-averaged RF power coupled to plasma by the whole ECT system, (d) the CTL and DTL maximum voltages in the ECT circuit powered by the amplifier 'D2'; the traces correspond to the maximum values in the running 0.1s time window, (e) the VSWR in the OTL of the amplifier 'D2'.

**Figure 22.** Combined high power operations of the JET load-tolerant ICRH systems during ELMy plasma: (a) auxiliary heating power injected into plasma by the ICRH and Neutral Beam systems (b) intensity of $D_\alpha$-line emission from plasma, (c) Radial Outer Gap between the separatrix and the antenna limiters, (d) maximum antenna voltages in the ECT, 3dB and ILA systems and (e) power coupled to plasma by the ECT, 3dB and ILA systems; note, that only one half of the ILA was operational during the pulse.

The observed complications of the ECT control during high power operations could be resolved by improvement of the real-time algorithms adjusting the lengths of the CTL and DTL trombones and by certain hardware changes specific to the JET installation. The latter implies replacement of the slow and inertial DC motors driving the moving sections of the CTL and DTL trombones with stepper motors allowing more accurate and immediate response to the error signals. Together with such an upgrade, the automatic tracking algorithm could be made more directly relevant to the voltage balance without seriously affecting the impedance matching. Indeed, it could be shown (e.g. compare figure 18 and figure 23) that the frequency error signal $E_F$ presently used to produce the $E_{CTL}$ and $E_{DTL}$ error signals on the basis of the formulae (20) could be replaced with the signal related to the ratio $E_V = \left|V_{CTL}^{max}\right| / \left|V_{DTL}^{max}\right|$ of the maximum voltages in the CTL and DTL lines. Numerical simulations confirm that such $E_V$ error signal (together with the existing stub error signal $E_S$) will provide more robust trombone length control being more sensitive to the voltage imbalance at low VSWR values and without compromising the capabilities of impedance matching. The proposed relatively simple modifications will facilitate optimisation of the ECT performance at high power levels removing the necessity for tedious 'manual' refinement of the trombone lengths.

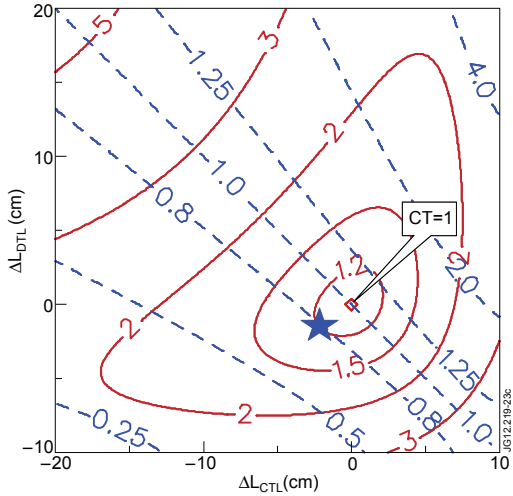

**Figure 23.** Dependence of the ratio of the maximum voltage amplitudes in the conjugated lines CTL and DTL on the deviation of the CTL and DTL lengths from their target values ensuring perfect matching: the dashed and the solid curves represent the $\left|V_{\text{CTL}}^{\max}\right|/\left|V_{\text{DTL}}^{\max}\right|$ ratio and the OTL VSWR respectively. The position of the star symbol roughly corresponds to the experimental conditions at the beginning of the RF pulse illustrated on figure 21.

## 6. Summary

The External Conjugate-T impedance matching system for two A2 ICRH antennas has been successfully integrated into the RF plant at JET. The system allows reliable uninterrupted injection of the RF power into H-mode plasmas in the presence of strong antenna loading perturbations during ELMs. Within the ECT, the corresponding current-carrying straps belonging to separate ICRH antennas (antennas 'C' and 'D') are paired to form complex conjugate impedances at coaxial T-junctions located outside the vacuum vessel. These parallel conjugate impedances are tuned using line stretchers (trombones) such that the resulting T-junction impedances have small and purely real values. These impedances are further transformed to the amplifier output impedance by the adjustable stub-trombone tuners.

The adopted ECT circuit configuration allows arbitrary phasing of the straps within the antennas and eliminates the negative influence of mutual coupling between the conjugated straps on the system performance. Dedicated arc detection techniques and real-time matching algorithms have been developed and implemented as a part of the ECT project. The new Advanced Wave Amplitude Comparison System (AWACS) operating in parallel with the traditional high-VSWR trip system has proven highly effective in the detection of arcs both between and during ELMs. The developed real-time control of the trombone lengths was found to be adequate for achieving good impedance matching during variable loading conditions; further efforts are required to automatically balance the voltages in the conjugated lines to ensure consistent high power operations. The implemented ECT scheme is compatible with the conventional configuration of the RF plant where the antennas are energized individually, enhancing the plant flexibility and facilitating the antenna conditioning procedures. Both plant operational modes share the control, protection and data acquisition electronics; the introduction of the ECT required minimal modification of the established infrastructure.

The ECT system has been fully commissioned at the frequencies of 32.5 MHz, 42.5 MHz, 46.0 MHz and 51.0 MHz answering the demands of most of the discharge scenarios used at JET. The ECT operational experience includes a large variety of antenna loading conditions such as vacuum, L-mode plasmas with strong 'sawtooth' activity, plasmas with L-H-mode transitions and ELMy H-mode plasmas with Radial Outer Gaps (i.e. the mid-plane limiter-separatrix distances) in the range of 4-14 cm. The experiments were performed mostly at standard $(0,\pi,0,\pi)$ phasing of the antenna straps and transition to alternative $\pm(0,\pi/2,\pi,3/2\pi)$ and $(0,\pi,\pi,0)$ phasing has been found straightforward. The circuit's high tolerance to ELMs predicted during simulations has been fully confirmed experimentally. Maximum trip-free time-average power levels coupled to ELMy H-mode plasmas using the ECT system have reached 4 MW; these levels strongly depend on peculiarities of JET discharge scenarios which define the antenna loading resistance between ELMs. In practice, the power is limited by the RF plant protection system that monitors the maximum voltages on the antenna straps and clamps the power as the voltages approach the breakdown threshold of 30-33 kV.

The improved RF plant performance during ELMy plasmas offered by the ECT system, together with the implementation of the 3dB couplers on another pair of A2 antennas and the installation of the ITER-like antenna, have considerably enhanced the research program at JET. The combined operations of all the ELM-tolerant systems in 2009 allowed delivery of more than 8 MW into H-mode plasmas. Such power capabilities made it possible to perform at JET a series of experiments requiring significant ICRH contribution to the total power balance in the ELMy discharges. In particular, a comparison between dominant ICRH and dominant NBI heated ELMy H-mode discharges was performed [49] providing valuable contribution to verification of ITER baseline scenarios. Availability of high RF power levels during ELMy discharges also helped to make substantial progress in the research of steady-state advanced scenarios [50]. The main experimental results of recent high-power ICRH operations at JET and the contribution of alternative ELM-tolerant ICRH systems are summarized in [11].

The development and successful implementation of the ECT system at JET expands the arsenal of technological tools available for reliable delivery of ICRH power into reactor-relevant plasmas and increases confidence in the viability of ICRH&CD in future fusion devices.

**Acknowledgments**

This work, supported by the European Communities under the contract of Association between EURATOM and CCFE, was carried out within the framework of the European Fusion Development Agreement. The views and opinions expressed herein do not necessarily reflect those of the European Commission. This work was also part-funded by the RCUK Energy Programme under grant EP/I501045.